\newcommand{\chisq}{\ensuremath{\chi ^2}}

\newcommand{\SNIa}{\ensuremath{\mathrm{SN~Ia}}}
\newcommand{\SNIae}{\ensuremath{\mathrm{SNe~Ia}}}
\newcommand{\degree}{\ensuremath{^\circ}}

\newcommand{\trf}{\ensuremath{~\tau~}}
\newcommand{\RGf}{\ensuremath{\mathrm{RG}_\mathrm{frac}~}}
\newcommand{\URF}{\ensuremath{\mathrm{U}~}}
\newcommand{\BRF}{\ensuremath{\mathrm{B}~}}
\newcommand{\VRF}{\ensuremath{\mathrm{V}~}}
\newcommand{\BmV}{\ensuremath{\mathrm{f}_\mathrm{B-V}~}} 
\newcommand{\UmV}{\ensuremath{\mathrm{f}_\mathrm{U-V}~}} 
\newcommand{\UmB}{\ensuremath{\mathrm{f}_\mathrm{U-B}~}} 
\newcommand{\pv}{\ensuremath{\mathrm{\emph{p}-value}}}

\documentclass{emulateapj}
\usepackage{graphicx}
\usepackage{natbib}
\usepackage{amssymb}

\begin{document}

\title{Constraining Type Ia Supernovae progenitors from three years of SNLS data.}
\author{
F.~B.~Bianco,\altaffilmark{1,2}
D.~A.~Howell,\altaffilmark{1,2}
M.~Sullivan,\altaffilmark{3}
A.~Conley,\altaffilmark{4}
D.~Kasen,\altaffilmark{5}
S.~Gonz\'{a}lez-Gait\'{a}n,\altaffilmark{6}
J.~Guy,\altaffilmark{7}
P.~Astier,\altaffilmark{7}
C.~Balland,\altaffilmark{7,8}
R.~G.~Carlberg,\altaffilmark{6}
D.~Fouchez,\altaffilmark{9}
N.~Fourmanoit,\altaffilmark{7}
D.~Hardin,\altaffilmark{7}
I.~Hook,\altaffilmark{3,10}
C.~Lidman,\altaffilmark{11}
R.~Pain,\altaffilmark{7}
N.~Palanque-Delabrouille,\altaffilmark{12}
S.~Perlmutter,\altaffilmark{13,5}
K.~M.~Perrett,\altaffilmark{6,14}
C.~J.~Pritchet,\altaffilmark{15}
N.~Regnault,\altaffilmark{7}
J.~Rich,\altaffilmark{12}
V.~Ruhlmann-Kleider\altaffilmark{12}
}

\altaffiltext{1}{Department of Physics, University of California Santa Barbara,
Mail Code 9530,  Santa Barbara CA 93106-9530}\email{fbianco@lcogt.net}
\altaffiltext{2}{Las Cumbres Observatory Global Telescope Network, Inc.
6740 Cortona Dr. Suite 102, Santa Barbara, CA 93117}
\altaffiltext{3}{Department of Physics (Astrophysics), University of Oxford, DWB, Keble Road, Oxford, OX1 3RH, UK}
\altaffiltext{4}{Center for Astrophysics and Space Astronomy, University
of Colorado, 593 UCB, Boulder, CO, 80309-0593, USA}
\altaffiltext{5}{Departments of Physics and Astronomy, University of California, Berkeley}
\altaffiltext{6}{Department of Astronomy and Astrophysics, University of Toronto, 50 St. George Street, Toronto, ON, M5S 3H4, Canada}
\altaffiltext{7}{LPNHE, Universit\'e Pierre et Marie Curie Paris 6, Universit\'e Paris Diderot Paris 7, CNRS-IN2P3, 4 Place Jussieu, 75252 Paris Cedex 05, France}
\altaffiltext{8}{Universit\'e Paris 11, Orsay, F-91405, France}
\altaffiltext{9}{CPPM, CNRS-IN2P3 and Universit\'e Aix-Marseille II,
Case 907, 13288 Marseille Cedex 9, France}
\altaffiltext{10}{INAF - Osservatorio Astronomico di Roma, via Frascati 33, 00040 Monteporzio (RM), Italy}
\altaffiltext{11}{Australian Astronomical Observatory, P.O. Box 296,
Epping, NSW 1710, Australia}
\altaffiltext{12}{CEA, Centre de Saclay, Irfu/SPP, F-91191
Gif-sur-Yvette, France}
\altaffiltext{13}{LBNL, 1 Cyclotron Rd, Berkeley, CA 9472}
\altaffiltext{14}{Network Information Operations, DRDC Ottawa, 3701 Carling Avenue, Ottawa, ON, K1A 0Z4, Canada}
\altaffiltext{15}{Department of Physics and Astronomy, University of
Victoria, PO Box 3055 STN CSC, Victoria BC, V8T 1M8, Canada}

\begin{abstract}
While it is generally accepted that Type Ia supernovae are the result of
the explosion of a carbon-oxygen White Dwarf accreting mass in a binary
system, the details of their genesis still elude us, and the nature of
the binary companion is uncertain. Kasen (2010) points out that the
presence of a non-degenerate companion in the progenitor system could
leave an observable trace: a flux excess in the early rise portion of
the lightcurve caused by the ejecta impact with the companion
itself. This excess would be observable only under favorable viewing
angles, and its intensity depends on the nature of the companion.  We
searched for the signature of a non-degenerate companion in three years
of Supernova Legacy Survey data by generating synthetic lightcurves
accounting for the effects of shocking and comparing true and synthetic
time series with Kolmogorov-Smirnov tests. Our most constraining result
comes from noting that the shocking effect is more prominent in
rest-frame B than V band: we rule out a contribution from white
dwarf--red giant binary systems to Type Ia supernova explosions greater
than 10\% at $2\sigma$, and than 20\% at $3\sigma$ level.

\end{abstract}

\section{Introduction}\label{sec:intro}
\setcounter{footnote}{0}
Type Ia supernova (\SNIa) explosions are marvelous astrophysical
tools, and they currently offer the most precise way of constraining
dark energy~\citep{Sullivan11}.
Today, several thousand \SNIae~have been observed and 
theoretical models and simulations are progressing rapidly (see for example \citealt{2010arXiv1008.2801A}), and many aspects of \SNIa~explosions can be reproduced in
detail. However, these cosmic explosions, studied for decades, still are not fully understood. Particularly, we lack a solid understanding of the progenitor systems.   
There is consensus that \SNIae~arise from the thermonuclear explosion of
a carbon-oxygen (C-O) white dwarf (WD), but is the WD in a binary system with a main sequence (MS) or red giant (RG)
star, accreting mass from the companion to approach the Chandrasekhar
limit (\emph{single degenerate} --SD-- scenario)? Or is the explosion
caused by the merger of two WDs in a compact binary system (\emph{double
degenerate} --DD-- scenario) ? Constraining the progenitor scenarios is
key for learning the details of \SNIa~explosion physics, and to improve
our understanding of the effects of environment on \SNIa~explosions and thus
of
the systematics that still affect the constraints on cosmology
derived from SN surveys (\citealt{2009ApJS..185...32K, Guy10},
\citealt{2007ApJ...666..694W}; for a
review of SN Ia cosmology see~\citealt{HowellReview}).

No progenitor system of a \SNIa~has yet been observed prior to explosion: these
binary systems would be very faint and undetectable, at this time, in
extra-galactic surveys. Population synthesis and environment studies
have not been able to firmly set constraints on the \SNIa~
progenitors. From the theoretical point of view, generating \SNIae~in the DD scenario presents difficulties. The mass transfer only successfully leads to a deflagration
if it occurs at a rate significantly slower than the Eddington limit, through the formation of a thick disk \citep{Nomoto85}, 
and even then fine tuning of various parameters might be needed (see~\citealt{2007ASPC..372..375T} for a brief review, and the references therein).  

Some observational evidence might already disfavor SD progenitors.
While the WD accretes mass from a companion in the SD scenario, the
system should emit X-ray radiation for an extended period of time. Under
the assumptions of continuous duty cycle, and that all SD progenitors
would emit in the X-ray, the \SNIa~rate  is too high by over an order
of magnitude compared to the number of X-ray sources observed in nearby
elliptical galaxies~\citep{2010Natur.463..924G}, as well as soft X-ray
sources in our own galaxy~\citep{2010ApJ...712..728D}. While this
evidence can be used to set upper limits to SD progenitors
,~\citet{2010ApJ...719..474D} points out that there are too few super
soft X-ray sources, sources with energy typically 10 to $100~\mathrm{eV}$ and luminosity $10^{-37}$ to $10^{-38}~\mathrm{erg/s}$, to account for the Type Ia rate within the DD
scenario as well, suggesting instead that super soft X-ray radiation may
not always be emitted in nuclear-burning white dwarf systems and that it
could be absorbed within the system itself. 
A few peculiar \SNIae~have been observed in the past few
years to produce $^{56}\mathrm{Ni}$ masses close to, or in excess of,
the $1.4\mathrm{M}_\odot$ theoretical limit for a WD mass: the
Chandrasekhar limit. These apparent super-Chandrasekhar \SNIae~may originate from the merger of two WDs, in the DD scenario~\citep{2006Natur.443..308H,2011MNRAS.410..585S,
2010ApJ...715.1338Y}.

\citet{Kasen09} -- hereinafter K10 -- explores the effect that a non-degenerate companion star would have on the observables of the explosion, and 
 shows that in the SD scenario, the presence of a companion may manifest
 itself in the early days after the SN explosion as a flux excess. As
 the cloud of SN ejecta expands, it collides with the companion. This
 impact shocks the expanding material creating a hole in the otherwise
 optically thick ejecta shell through which radiation can escape. An
 excess flux is produced in the shocked gas, propagating in the
 direction of the observer, and it should be detectable when the
 geometry is favorable and the observer looks into the companion star.
As the the equilibrium
temperature of the shocked debris is inversely proportional to the
distance from the WD center to the power of 3/4 (see Eq. 7 in K10), the flux excess is larger the larger
the separation from the companion, and assuming Roche lobe overflow,  RG companions will leave the most prominent signature. The intensity of the feature is higher in bluer bands (see Section~\ref{sec:models}). The effect only lasts a few days, completely vanishing by 10 days after explosion.

With detailed early time lightcurves we may be able to identify the
progenitor system of a particular SN explosion, singling out events
generated by red giant progenitors when seen from favorable
angles. Unfortunately, early detailed SN lightcurves, with daily or so cadence,
are still rare.  New surveys like the PanSTARRS Medium Deep Survey (PS1,
\citealt{2010ApJ...724L..16P}) and the Palomar Transient Factory (PTF,
\citealt{2009PASP..121.1395L}) provide well sampled early \SNIa~lightcurves, which might lead to the identification of progenitors in
individual cases, as might early UV follow up.

With the large collection of lightcurves provided by surveys such as the Supernova Legacy Survey (SNLS,~\citealt{2006A&A...447...31A}) and the Sloan Digital Sky Survey (SDSS, \citealt{2009ApJS..182..543A}) the companion scenarios can be constrained in a statistical fashion.
The SDSS collaboration recently searched for evidence of an early flux
excess due to shocking in high signal-to-noise ratio (S/N)
spectroscopically confirmed \SNIa~lightcurves from the SDSS-II
survey. Finding no evidence of shocking-related excess in a subset of
108 confirmed \SNIae~with well observed early time behavior,
~\citealt{2010sdss} conclude that RG's cannot be the main channel for
\SNIa~explosions. Here we use confirmed \SNIae~from the first three years of SNLS data to set an upper limit to the contribution of RGs to \SNIa~progenitors. 

After describing the dataset used here, consisting of spectroscopically confirmed \SNIae~from the first three years of SNLS, and the processes used to standardize the data and generate composite lightcurves (Section~\ref{sec:data}), in Section \ref{sec:models} we summarize the results presented in K10 and show our rendering of these models. We then describe the statistical tests that allowed us to derive constraints to the contribution from RG progenitors to \SNIa~explosions (Section~\ref{sec:tests}).
We also extended our analysis beyond the spectroscopically confirmed
sample to include photometrically selected \SNIa~lightcurves, showing
that lightcurves affected by shock were not rejected as \SNIa~spectroscopic follow up candidates in SNLS because of a selection bias,
and that our conclusions extend to the photometrically selected
\SNIae~(Section~\ref{sec:unconf}). Finally, in Section \ref{sec:conc} we summarize our conclusions and outline future work.

\section{SNLS data, third year}\label{sec:data}
The dataset used here is described in detail in
\citet{2011ApJS..192....1C}, \citet{Guy10}, \citet{Santi10} and
\citet{Bazin10}. We use data from the first 3 years of the SNLS. The SNLS is
a rolling survey that gathered photometric data at the
Canada-France-Hawaii-Telescope (CFHT). Two independent photometric
pipelines, based in France and Canada, are used for the SNLS data
reduction \citep{Bazin10, 2010AJ....140..518P}. Here we use the photometry output of the
French pipeline.  SNLS lightcurves, originally collected
in \emph{griz} 
\citep{Regnault10}, are \emph{k-}corrected
\citep{2007ApJ...663.1187H}, and standardized by applying a stretch
factor, to broaden or narrow the rest-frame timescale of the
lightcurve~\citep{1997ApJ...483..565P}, in order to generate rest-frame
\BRF and \VRF band lightcurves\footnote{Note that this is different from
what is done in the processing of SNIa lightcurves for cosmology, where
the stretch correction is applied to the rest-frame template, to match
the data~\citep{2011ApJS..192....1C, Sullivan11}.}. We define the
variable \trf as
\begin{equation} 
\trf = \frac{t-t_\mathrm{max}}{s (1+z)},
\end{equation} where $t_\mathrm{max}$ is the date of maximum flux (in rest-frame \BRF
filter band), $z$ is the SN redshift and $s$ the stretch; $\trf$
represents the rest-frame, stretch-corrected, time to peak \BRF luminosity. 
We processed the lightcurves using the SiFTO method~\citep{Conley08} and
we used a single template 
to fit the data and stretch correct the lightcurves. 
In processing the SN data  we assumed a rise time $\tau_r=17.4$~days, the time elapsed between explosion and maximum B luminosity,
 according to what \citet{Santi10} finds in a similar (but larger) SNLS
 dataset, and it is also consistent with the rise time derived in
 \citet{2009sdss} from  SDSS-II \SNIae~($\tau_r=17.38\pm 0.17~\mathrm{days}$).

Our primary analysis is focused on spectroscopically confirmed \SNIae. 
The first 3 years of SNLS data offer over 200 spectroscopically
confirmed \SNIae~lightcurves. The original dataset was reduced to 87 SN
lightcurves by applying quality and redshift cuts described below.
The final dataset uses only SNe that satisfy the following requirements:

\begin{itemize}
\item spectroscopically confirmed Type Ia SNe at redshift $z < 0.7$ (135 lightcurves)
\item the total reduced \chisq~for the SiFTO template fit, applied to
      epochs  $\trf > -10$ days, is better than 3.0 (130 lightcurves)
\item the error in the determination of the peak date is $\Delta d_\mathrm{max}<0.7$ days (117 lightcurves)
\item have at least three data points in rest-frame \BRF and three data points in rest-frame \VRF band  in the rise portion of the lightcurve, $-10 \leq \trf \leq0$~days, to ensure the quality of the pre-peak fit (87 lightcurves).
\end{itemize}

The excess due to shocking may be visible up to 10 days after explosion
(K10), or $\trf \sim -8$ given our choice of $\tau_r=17.4$~days. No data
prior to $\trf=-10$ days were used to standardize the data and generate
our composite lightcurves in order to avoid including in the lightcurve
fitting process data points potentially affected by the excess that we are seeking.  
Different choices of minimum day (between -10 and -7) were also tested and they do not  affect our result. However removing points earlier than $\trf = -8$ causes, in a few cases, a poor lightcurve fit, and thus a larger scatter in the data. Visual inspection reveals that none of those \SNIae~for which the fit parameters significantly change if data points between $\trf = -7$ and $-10$ are excluded is actually affected by shocking. Thus we conclude that including points at $-8 < \trf < -10$ only strengthens the significance of our results. 

The templates adopted to process the data ({\tt Conley09c} and
{\tt Conley09f}\footnote{We see no difference
in our results choosing either {\tt Conley09c} or
{\tt Conley09f}, and where not specified we will refer
to {\tt Conley09f} throughout the paper.}, \citealt{Conley08})
assume a parabolic behavior in time prior to $\trf=-10$ days\footnote{In fact, the SiFTO method allows as well for a cubic fit to the early rise portion of the lightcurves, but this was found not to improve the lightcurve fit in most cases~\citep{Conley08}.}, as described in~\citet{2001ApJ...558..359G}, \citet{Conley08}, and in the references therein
: 
\begin{equation}
f(t)=\alpha (\tau-\tau_r)^2,
\end{equation}
where $f$ is the flux as a function of time $t$, and rise time
$\tau_r$. This is consistent with a simple \emph{expanding fire ball}
modeling of the exploding ejecta (see for
example~\citealt{1999AJ....118.2675R}), with $\alpha$ representing the
rise ``speed,'' 
 which is what we expect in absence of shocking by a companion. We
refer the reader to  \citet{Santi10} for a detailed study of the
rise behavior of \SNIae~in the SNLS data.

\citet{2009sdss} found that a better fit to the SDSS-II data can be achieved using two separate templates to fit the rise and fall portion of the lightcurves, thus obtaining two stretch values. While using 2 stretches slightly improves the \chisq~per degree of freedom the individual lightcurve fits, F-tests show that for these SNLS data the improvement achieved using 2 stretches is not significant (see also~\citealt{Santi10}).  Furthermore, because we use only data points at $\trf \geq -10$, with the 5 day cadence of the SNLS data, we generally have only 2-3 points in the rise portion of each lightcurve that would be used for fitting. Fitting separately the rise and fall portions of the lightcurve then  exposes us to the risk of misfitting or over-fitting the rise portion. We conclude it is best to use a single stretch template  for the purpose of this analysis.

A more detailed description of the standard SNLS lightcurve processing can be found in \citet{2011ApJS..192....1C},  \citet{Guy10}, and \citet{2006AJ....132.1707C}. For a discussion on the SNLS photometric calibration see~\citet{2009yCat..35060999R}.

\begin{figure*}
\centerline{\includegraphics[width=0.48\textwidth]{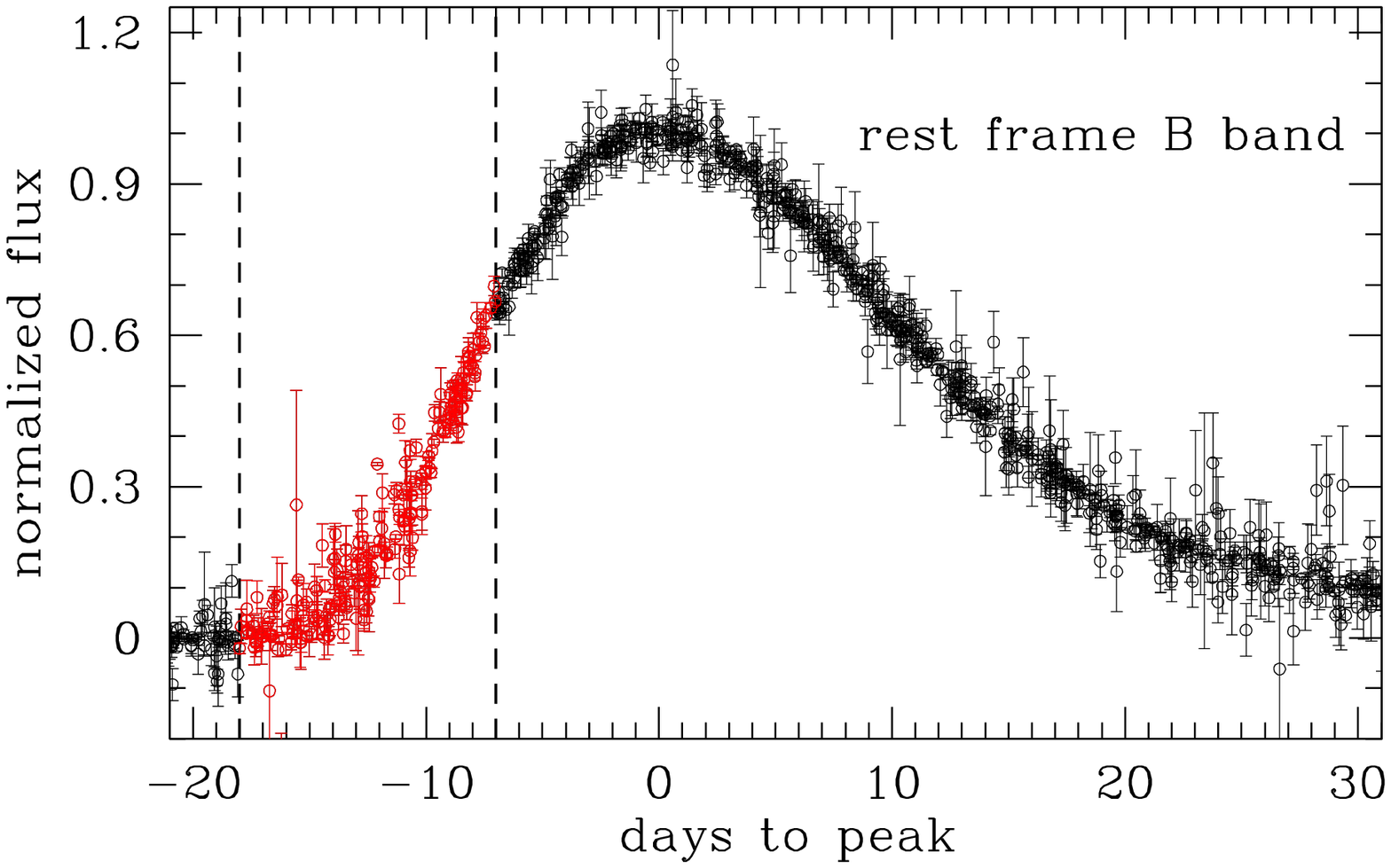}\includegraphics[width=0.48\textwidth]{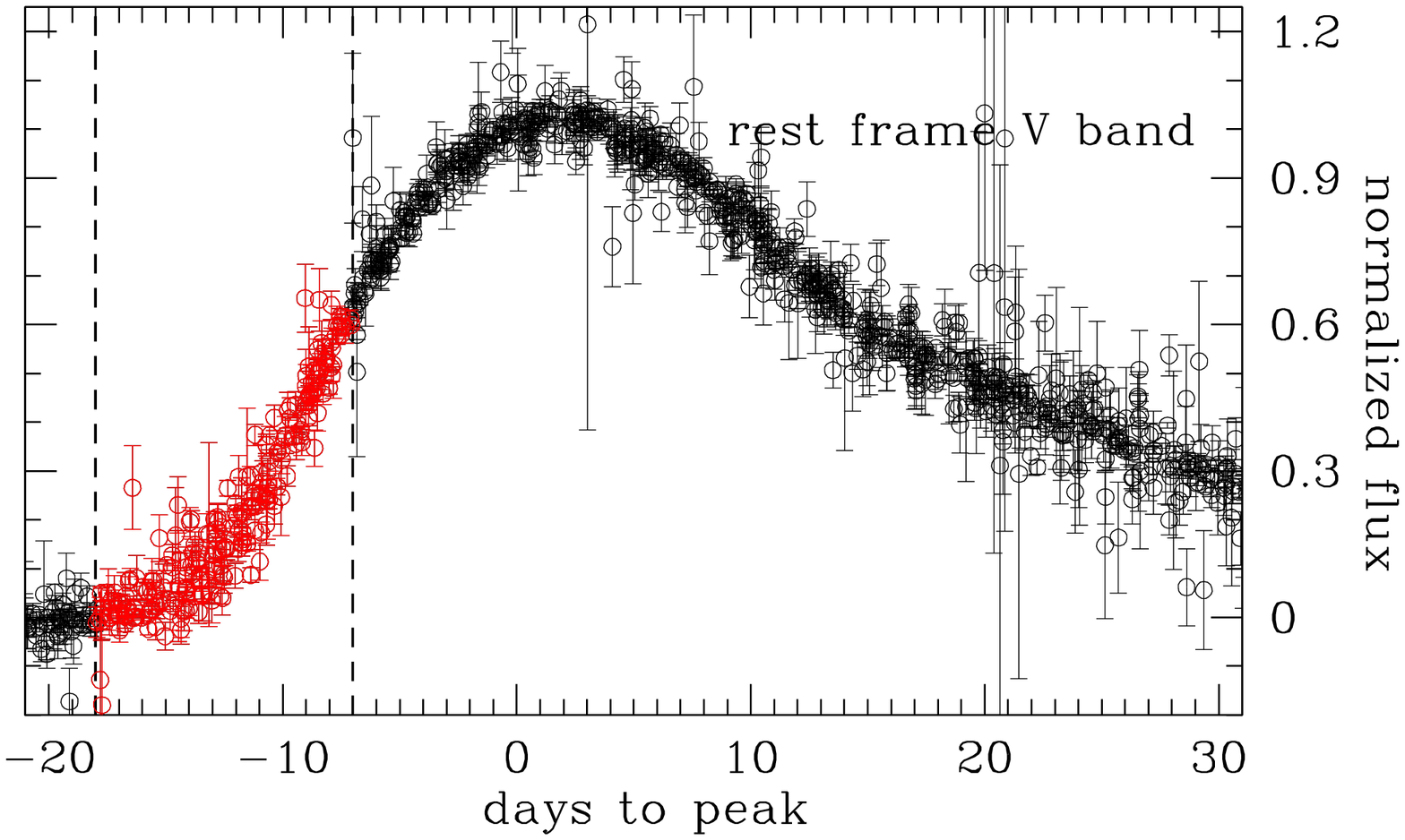}}
\caption{Composite rest-frame \BRF (left) and \VRF (right) SNLS lightcurves. Vertical lines delimit the region that could be affected by shocking, and the points within are plotted in red. Those are the data points potentially susceptible to the excess induced by the presence of a companion shocking the ejecta. }
\label{fig:complc}
\end{figure*}

The SNLS lightcurves, normalized to peak flux $f_\mathrm{peak}=1$ in each color channel, stretched, and \emph{k-}corrected as
described above, can be combined into a composite lightcurve: our
composite rest-frame \BRF (V) lightcurve contains a total of 1059 (1125)
data points between $\trf=-20 $ and $\trf=40$, and 202 (217) in the 10
days after explosion that would be affected by the flux excess: $-17.4
\leq \trf \leq -7.4$~days. The composite \BRF and \VRF lightcurves are
shown in Figure~\ref{fig:complc}, \BRF band flux on the left hand side
and \VRF band flux on the right hand side\footnote{For a discussion on the distinction between
photon-based, and energy-based flux see~\citet{Nugent02}. Throughout
the paper we refer to flux as photon-based flux.}. The data points potentially affected by the shocking excess are plotted in red, and included within vertical lines.

\section{Models}\label{sec:models}
After an explosion, the SN ejecta expands and collides with the
 companion star, if one exists. K10 showed  
 that this impact shocks the SN ejecta creating a hole in the expanding
material. Radiation can now escape from the otherwise optically thick
ejecta shell. An early X-ray emission, analogous to the X-ray flash in
core-collapse SN \citep{2008Natur.453..469S, 2009ApJ...702..226M}, should last minutes to hours, with little chance to be
observed. In UV and optical bands the flux excess lasts longer: the  gas
begins expanding to refill the hole carved by the companion. Radiation
 continues to diffuse out of this hot, shocked region, and it is
 observable until the $^{56}\mathrm{Ni}$ luminosity begins to dominate the lightcurve. The time scale for this process is roughly 5-10 days. The effect is more prominent in bluer bands and can span over an order of magnitude in flux in UV, generating an early peak even brighter than the peak luminosity in absence of shocking,  while it is substantially dimmed in V band. 

The size of the hole and of the shocked gas region, and thus the amount of
radiation escaping, depend
essentially on the solid angle subtended by the companion carving the
hole, and on velocity of the ejecta at the time of impact. 
The models assume that the
companion star is in Roche lobe overflow and that the exploding WD has
reached the Chandrasekhar limit. Under such assumptions these are
determined by the distance to the companion, and the geometry of
the system (semi-major axis of the binary orbit and size of the Roche lobe) is set by the \emph{nature} of the companion. This excess radiation is then a powerful tool to identify the SN progenitor system. 

\begin{figure*}
\centerline{\includegraphics[width=0.3\textwidth]{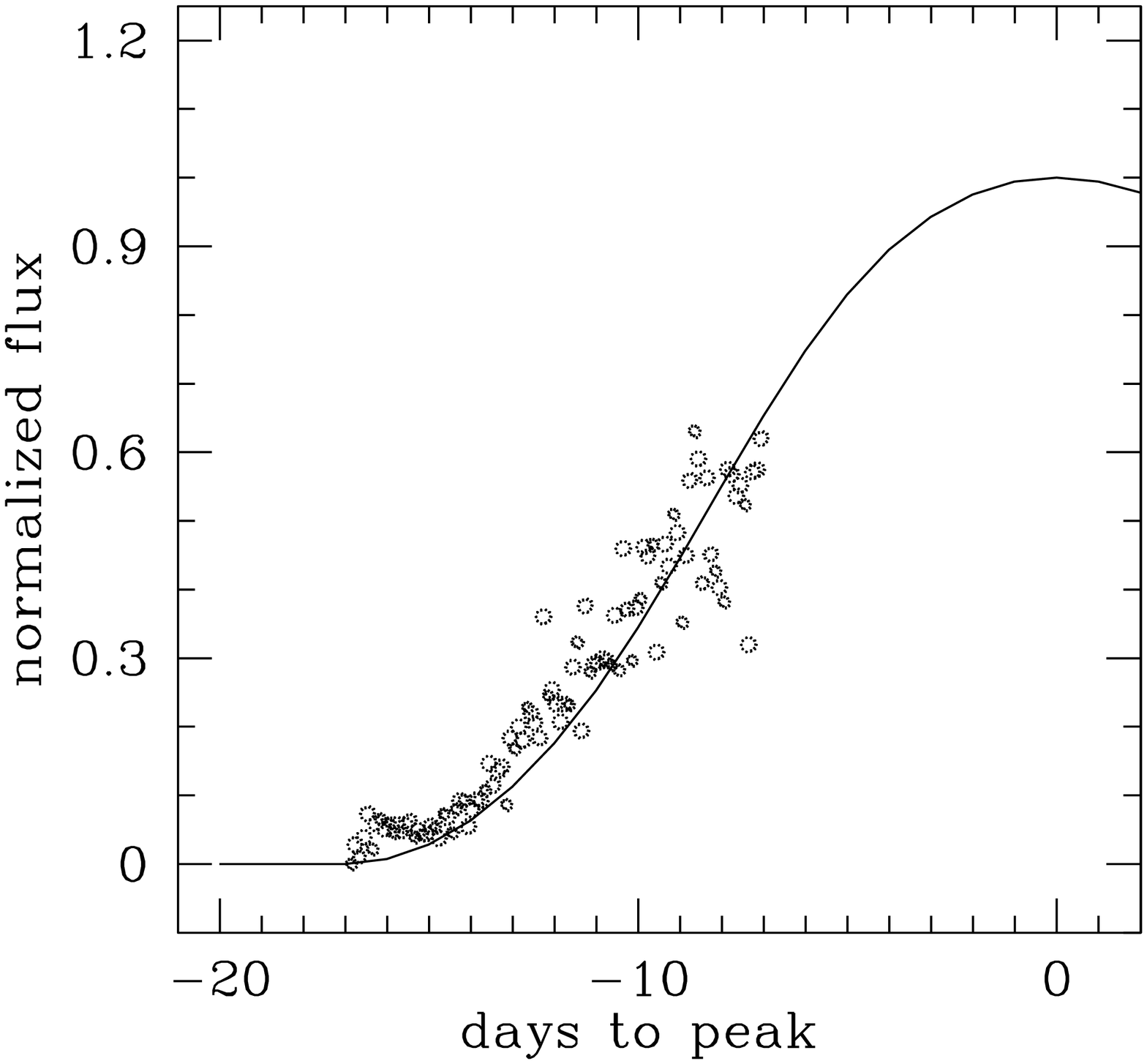}\includegraphics[width=0.3\textwidth]{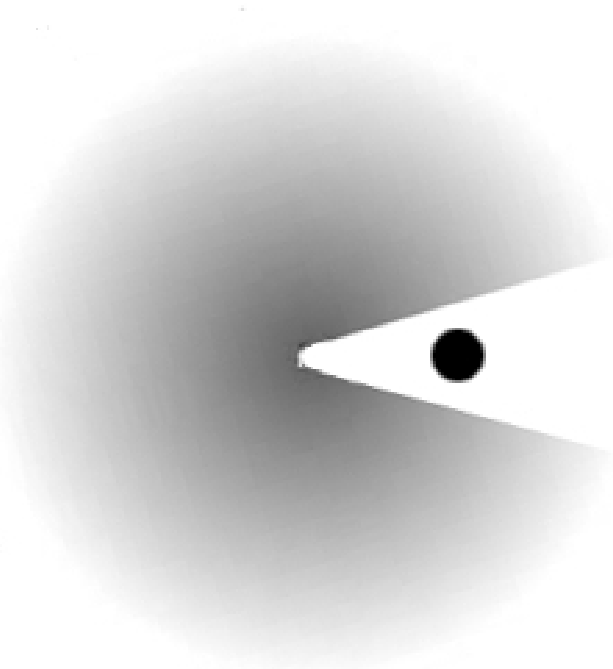}\includegraphics[width=0.3\textwidth]{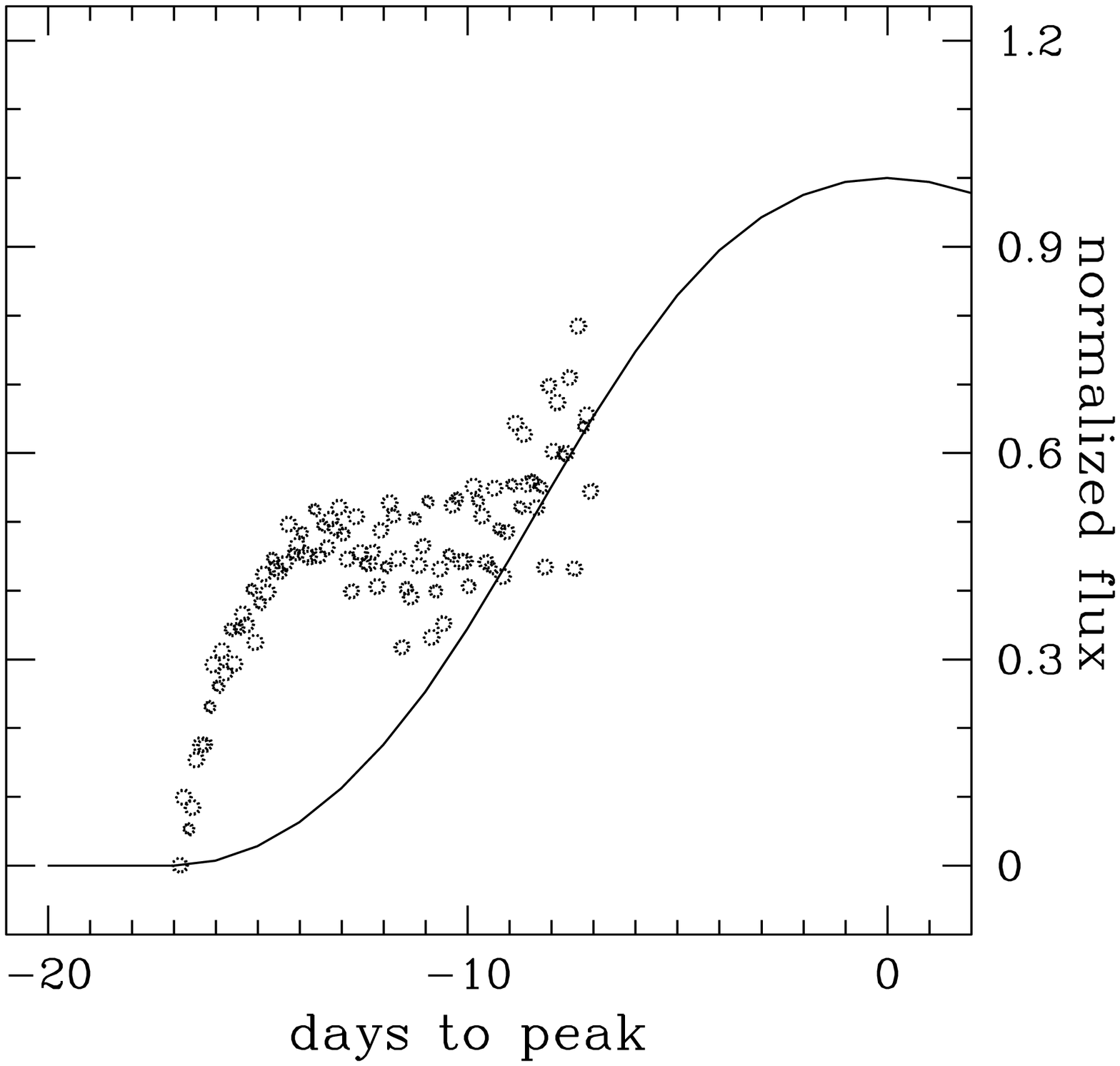}}
\caption{The K10 model for a WD accreting from a RG companion is shown. At the center is a schematic representation of the SD explosion scenario: in the expanding ejecta, gray, the impact with the companion star (black circle) has created a hole, here simplistically represented by a cone.  To the left and right, according to the corresponding view point, are the rise lightcurves for, respectively,  an observer looking in opposite direction from the companion (no excess), and looking into the companion and the hole created by the impact (maximum excess), for the case of a WD-RG progenitor system.
The scatter in the model is simply due to statistical noise in the numerical simulations (see Section~\ref{sec:models}). The solid line is the {\tt Conley09f} template.}
\label{fig:models}\end{figure*}

K10 considered 3 types of companions: a $2~M_\odot$ MS star,
at a distance $a=5\times 10^{11}~\mathrm{cm}$ from the WD, a $6~M_\odot$
MS companion at $a=2\times 10^{12}~\mathrm{cm}$, and a $1~M_\odot$ RG companion at $a=2 \times 10^{13}~\mathrm{cm}$. 
The effect is most prominent for observers viewing directly into, or at
small angles from, the companion (i.e. into the hole), or roughly $10\%$
of the time (Figure~\ref{fig:models}). However, particularly in the case
of  RG companions, where the excess is maximum, some
flux excess is observable even at large angles. We expect the
distribution of viewing angles to be uniform in nature, and in our set
of lightcurves. The presence of a flux excess may increase the
detectability of a SN, thus the orientation of the WD-companion system
may constitute a selection bias when operating close to a survey
detection limit. The cut at $z<0.7$ assures that we keep safely away
from the detection limit of SNLS. Furthermore the maximum excess in B
band is significantly smaller (a factor $>2$) than the peak luminosity,
and since SNLS is a rolling survey, covering each field every five days,
each lightcurve should, in principle, contain $\gtrsim 3$ data points
within 10 days of peak that would be brighter than the maximum shocking
flux. The shock-induced flux excess, therefore, does not significantly
increase the detectability of a SN in our sample, and we can assume that
the distribution of angles in our data is unbiased. Later we will expand
our analysis to a photometrically selected sample, to assess weather an early flux excess might have caused the
\SNIae~to be misclassified as a non-\SNIa, and thus not followed spectroscopically~(Section~\ref{sec:unconf}).

 The models are generated from  2-dimensional Monte Carlo radiation
 transport simulations that are subject to random sampling errors.  Such
 errors are purely statistical, and don't take into account any of the
 possible systematic errors or uncertainties in the model
 calculations. The statistical errors are accounted for throughout our analysis.

\begin{figure*}
\centerline{\includegraphics[width=0.48\textwidth]{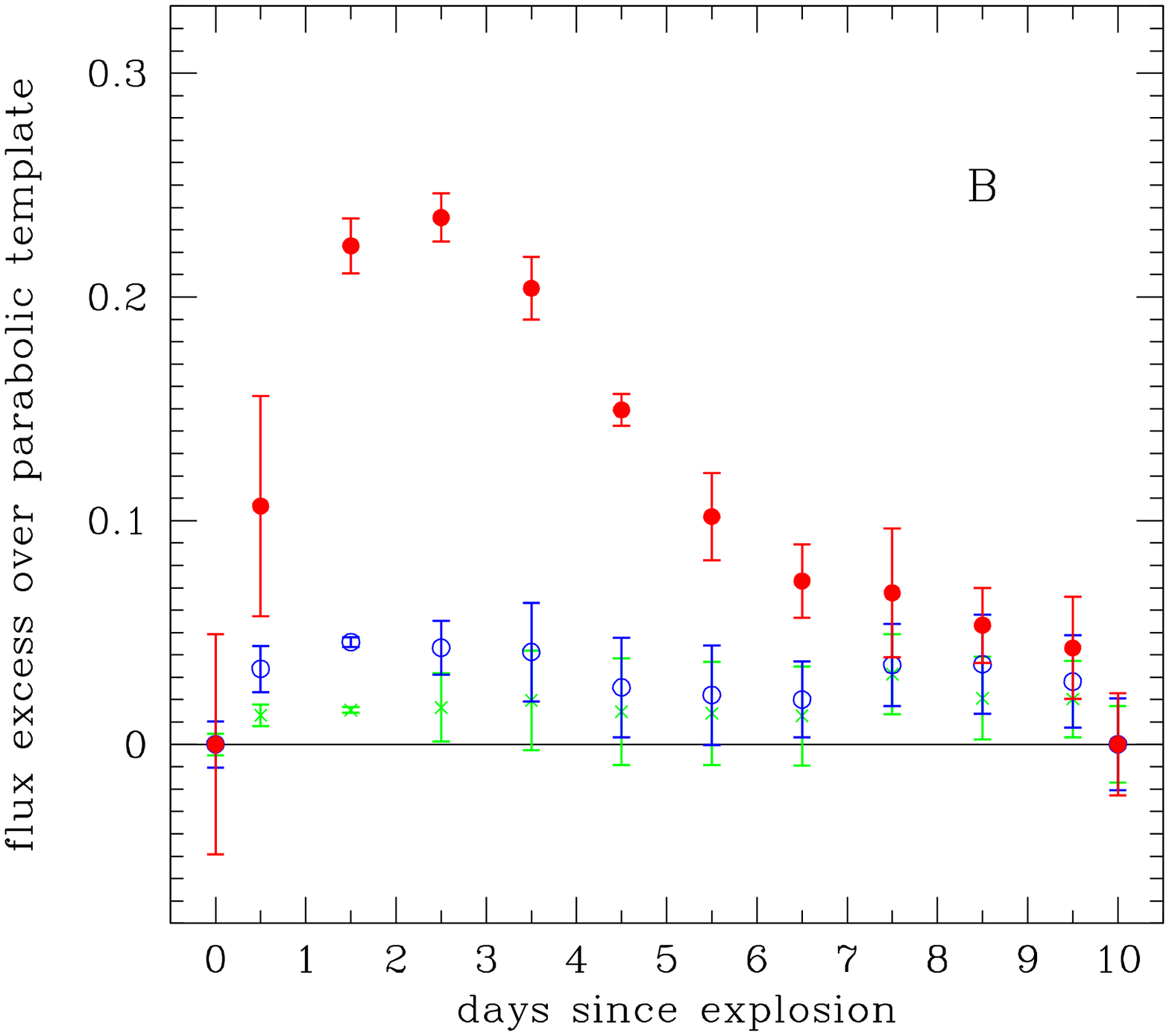}\includegraphics[width=0.48\textwidth]{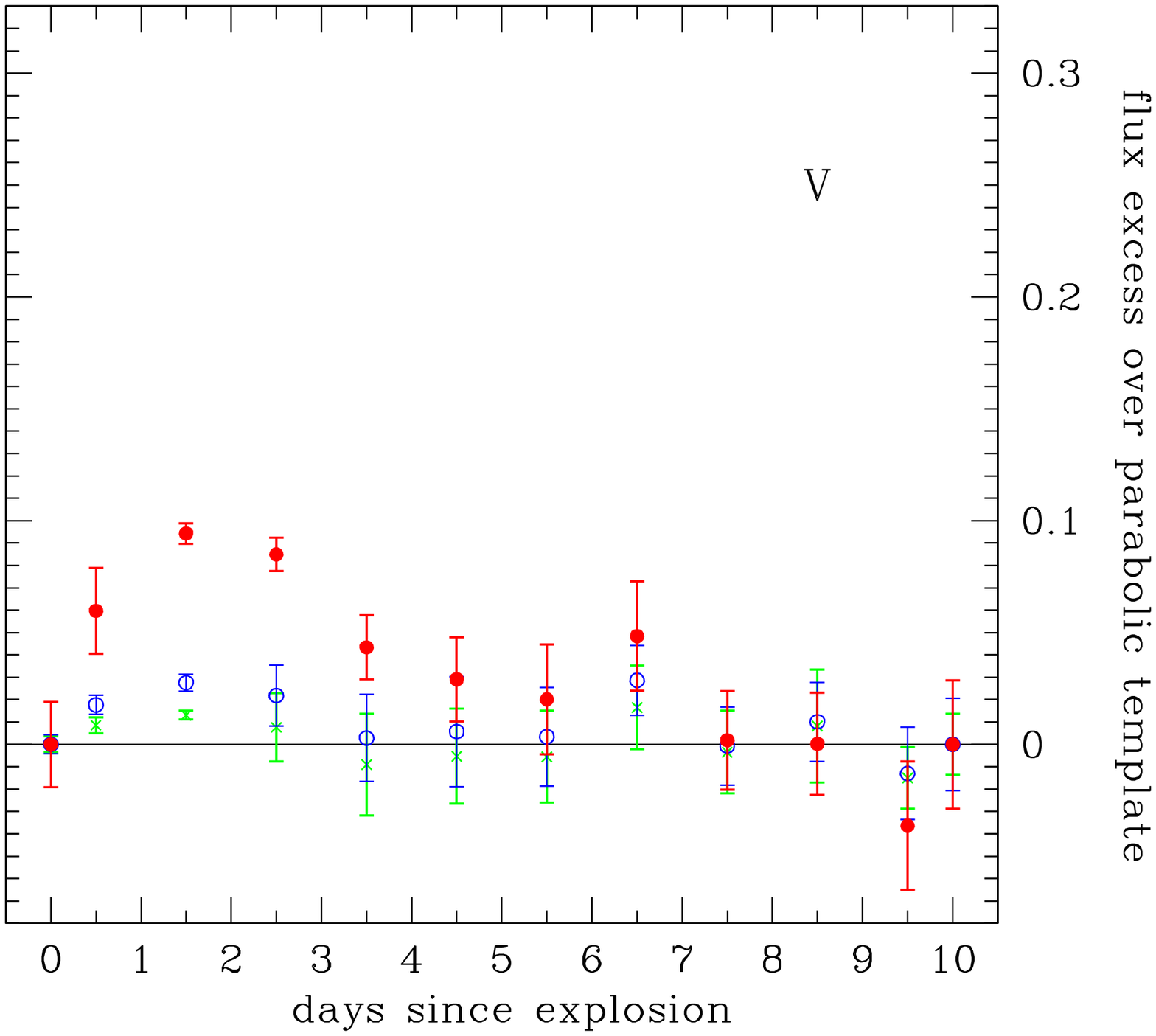}}
\caption{Models of excess emission over a nominal -- \emph{parabolic} --
 SN lightcurve template, in units of peak luminosity, signature of shocking by a companion star, for
 the cases considered in K10: a RG companion (red
 filled circles), a $6~M_\odot$ (blue empty circles) and a $2~M_\odot$ (green crosses) MS companion, all in Roche lobe overflow (separation from the core of the explosion $a~=~5 \times 10^{11},~2 \times 10^{12},~\mathrm{and~} 2 \times 10^{13}~$ cm respectively). The effect is shown averaged over all observing angles. \SNIa~spectra are generated from K10's simulations as described in Section~\ref{sec:manipul}, and filtered through standard B and V filters to generate the theoretical lightcurves. The error bars represent the scatter -- standard deviation --  in the models.
The left plot shows the effect in B and the right hand plot in V band. }
\label{fig:excessBV}
\end{figure*}

The excess generated by shocking is shown in Figure~\ref{fig:excessBV}, averaged over viewing angles, 
for all three progenitor scenarios in both B and V band.
It is evident that, while the RG progenitors generate a significant
excess, and a very distinct effect in both B and V band, the time
behavior for MS stars is only marginally changed in the presence of shocking, especially after averaging over viewing angles. Such a small deviation from the parabolic early rise behavior would hardly be detectable in the presence of the typical noise of SNLS data.
Therefore we restrict ourselves to the RG scenario and only try to constrain the RG contribution to SN progenitor systems. 
We also expect that the explosion in a DD scenario would show even
smaller, or no, deviations from a parabolic behavior. We thus tentatively associate the DD scenario to the standard template. Note, however, that it is possible, as mentioned in K10 and shown in \citet{Fryer10}, that in a WD merger gas would be blown out to large radii ($\sim 10^{13}~\mathrm{cm}$), producing a shock signature, with a UV excess possibly propagating through visible wavelengths. However, the simulations in \citet{Fryer10} generally produce lightcurves rather dissimilar from standard \SNIae, with broader visible band lightcurve, unlikely to match \SNIae~in our sample.

Where needed, we will assume that any explosion not generated by a RG-WD
binary pair has equal probability of arising from any of the three remaining scenarios.  

\subsection{Rendering of the models}\label{sec:manipul}
The K10 simulations generate full spectra of the SN
explosion, including the effects of shocking, at time intervals of 0.1
days 
for the first 10 days after explosion. The spectra are integrated on a day time scale and filtered through the same  V and B filters into which the SNLS data have been converted. This is done for every angular separation between the WD and the companion, in 40 equal intervals of observing probability, for the three progenitor scenarios considered: RG, $6~\mathrm{M}_\odot$ and $2~\mathrm{M}_\odot$ MS sub-giants. 

The K10 spectra are designed to reproduce the excess
generated by shocking \emph{on top of} a nominal template. The input
template in the models is irrelevant to the shocking physics. We use the
spectra for the smallest companion scenario ($M=2~M_\odot$) at the
largest angular separation ($\sim 180\degree$), where we expect the
effects of shocking to be entirely negligible, as a \emph{neutral}
template: the template in absence of shocking. To better reproduce what
we actually expect the result of shocking to look like in a SNLS
lightcurve we subtract the neutral template from the lightcurve
templates generated as described above. The new lightcurves are shown in
Figure~\ref{fig:excessBV}, averaged over all angles, and they describe
the excess due to shocking.  This excess can be added to the parabolic
portion of the  templates to reproduce what we would expect to see in our data. The reader is reminded that this portion of the lightcurve is \emph{not} used to standardize the SNLS data and generate the composite lightcurves. 

We now have template lightcurves for the first 10 days of a \SNIa~explosion in the SD scenario, with different companion stars and at different observing angles, which can be compared to the SNLS observations.

\section{Tests}\label{sec:tests}
\subsection{Template goodness of fit}\label{sec:gof}
We begin by noticing that the fit of the composite SN to the SN template (here {\tt Conley09c} and {\tt Conley09f} were used, with no significant differences) is worst in the region of interest for the shocking effect, $\trf < -7$ days to peak, in both \BRF and \VRF bands. 

\begin{figure*}[t]]
\centerline{\includegraphics[width=0.48\textwidth]{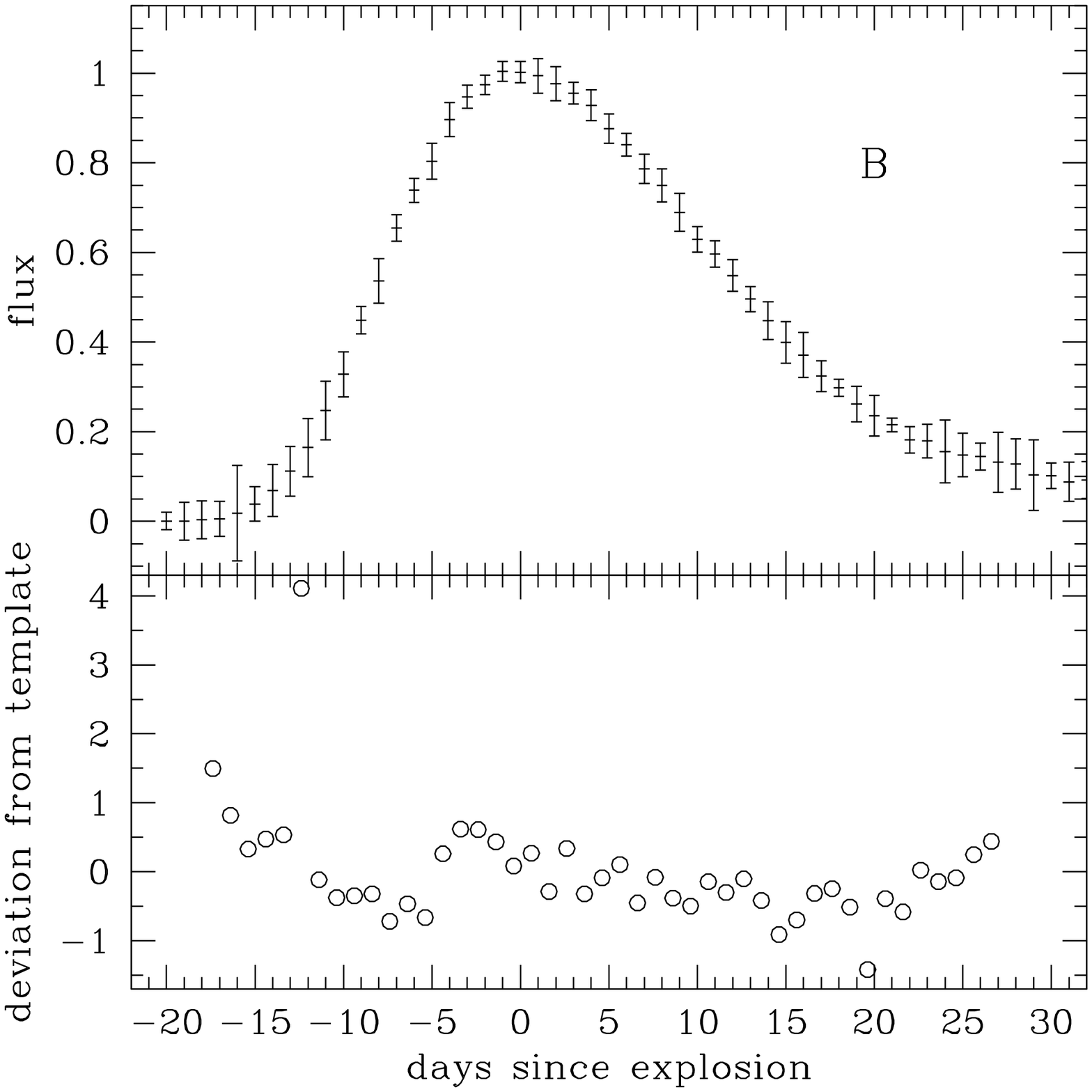}\includegraphics[width=0.48\textwidth]{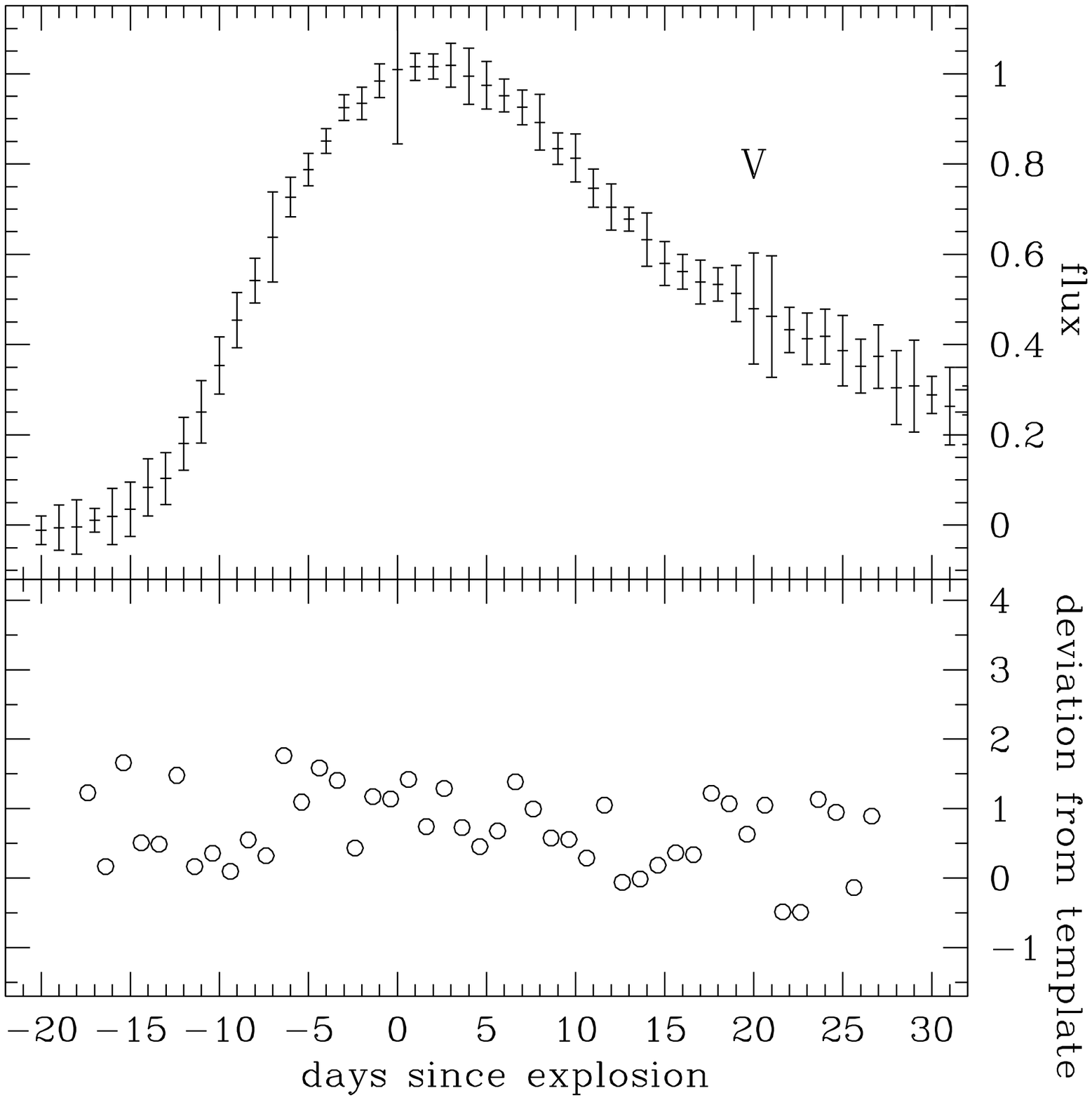}}
\caption{Top: median of the composite lightcurve shown in
 Figure~\ref{fig:complc} in rest-frame \BRF (left) and \VRF (right). The
 error bars represent the scatter in the measurements in the composite
 lightcurve. Bottom: a measure of the deviation of the median plotted in
 the top panels from the nominal SN templates, averaged day by day: $\sum_{t \mathrm{~in ~day}}^{~}\frac{f(t)-T(t)}{\sigma(t)}.$
The deviation of the composite lightcurves is most significant before
 \trf=-11 days in \BRF and before \trf=-7 days in \VRF.}
\label{fig:chi2}
\end{figure*}
Figure~\ref{fig:chi2} shows the median, binned by day, of the composite lightcurves (top
plots). The error bars represent the scatter -- standard deviation -- of
the individual measurements (the standard deviation is measured as
conventionally done with respect to the mean and we ignore the error
of  each measurement). 
The bottom plots show the deviation from the template as the  
difference between the data, $f$, and in the template,
$T$, at time $t$, over the error in the data $\sigma(t)$, averaged
over each day:
\begin{equation}
\sum_{t \mathrm{~in ~day}}^{~}\frac{f(t)-T(t)}{\sigma(t)},
\end{equation}\label{eq:deviation}
as an estimator of the goodness of fit of the template to the composite lightcurve.   
The propagation of errors along the time dimension is also
ignored.

The most significant deviation from the template happens in both \BRF
 and \VRF band roughly prior to  $\trf=-10$ days, with a clear excess in \BRF band at  $\trf\leq-12$ or in the first few days after explosion. It is intriguing that the deviation is more prominent in \BRF than in \VRF band, consistent with the chromatically biased effect that shocking by a companion would produce.
 Note, however, that this is the portion of the lightcurve that is not
 fit to the template, and it is thus not surprising to see a larger
 scatter here. 
\subsection{Simulations}\label{sec:sim}
Having found a deviation from our fiducial \SNIa~template in the early days
 after explosion, we test if this can be attributed to shocking by a
 companion.

With the K10 templates in hand we can create synthetic
\SNIa~rise lightcurves that incorporate the effect of shocking for the
different progenitor scenarios, as seen from different viewing angle. 
We start off with the standard parabolic rise templates ({\tt
Conley09f}). In each band we add the excess described by the models
(Figure~\ref{fig:excessBV}) to our standard template. For each epoch
corresponding to the SNLS data, we draw a data point from
the new template thus obtained. To choose the viewing angle from which this
data point should come we draw angles with equal probability
between 0 and 180\degree. 

Families of synthetic lightcurves are generated using increasing
contributions of data points from the RG progenitor scenario, and drawing
the remaining points equally likely from a parabolic template (DD
scenario), the $2~M_\odot$ and the $6~M_\odot$ MS progenitor
scenarios. We generate families with 0\% RG contribution, to 100\% RG
contribution, in steps of 10\%. A finer grid was also tested, but a
resolution of 10\% in the RG contribution is adequate to represent the
progenitor populations given our errors.

In our test we have to account for  the errors in both the templates and
the data. Thus at a given epoch, in generating the simulated data
points, we draw each flux value from a Gaussian distribution around
the template value at the corresponding epoch, where the width of the Gaussian is the sum in quadrature of the errors in the SNLS data and in the model.

For each \RGf between $\RGf=0\%$ and 100\% we create 100 sets of
simulated observations, in steps of 10\%, thus generating 100 synthetic
lightcurves per \RGf, each one the size of the rise portion of the lightcurve: 202 points in B and 217 in V band.

\subsection{One band K-S test}\label{sec:onebandks}

\begin{figure}[h]
\centerline{\includegraphics[width=0.48\textwidth]{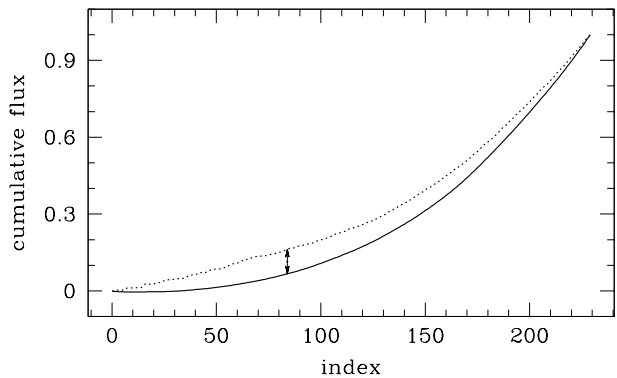}}
\caption{Cumulative, normalized flux distribution for the rise portion of the SNLS composite \BRF band lightcurve (solid line) and for one of the synthetic realizations (dotted line), with a contribution of RG progenitors $\RGf=30\%$ (see Section~\ref{sec:sim}). The data points in the lightcurves are ordered by flux and the $x$-axis is simply the rank of each sorted data point. The maximum vertical deviation between the true and synthetic cumulative flux distributions is marked by an arrow. From this value we asses the probability that the true data and this synthetic realization came from the same original distribution.}
\label{fig:cumulative}
\end{figure}

We then compare each population of synthetic lightcurves to our
composite lightcurves.  We chose the non-parametric 2-sample
Kolmogorov-Smirnov (K-S) test~\citep{1983MNRAS.202..615P} to do so. This
simple statistical test measures the maximum distance between the
simulated and true cumulative lightcurves, and we find it is a sensible
statistic to determine the presence of an excess in data with large
scatter. The K-S tests are non parametric, and thus do not require us to
make any assumptions on the distribution of data, and, unlike for
example the Pearson's $\chi^2$ statistics~\citep{Rice01}, do not require
binning the data.  The SNLS data points are sorted by flux, and added to
generate a cumulative flux distribution. Similarly, a cumulative flux
distribution is generated for each synthetic realization. The maximum
distance between two cumulative distributions is a measure of the level
at which the null hypothesis that the data sets being compared come from
the same distribution, can be rejected (see
Figure~\ref{fig:cumulative}).  In other words: comparing the true data
with synthetic data generated from scenarios with different RG
contributions, we test if the data comes from a distribution with a
given fraction of RG progenitors, $\mathrm{RG}_\mathrm{frac}$.

For each $\mathrm{RG}_\mathrm{frac}$, we pair each of the 100 sets of simulated observations with our true lightcurve, apply a 2-sample K-S tests, and average over the 100 K-S numbers. This allows us to obtain the confidence level for the rejection of the null hypothesis -- that the two sets being compared come from the same distribution -- and its statistical errors, accounting for all sources of noise: the uncertainty in the models, in the data, and for the presumed diversity in \SNIa~progenitors.

Using only the \BRF data, and within $1\sigma$ error bars, we reject at roughly 90\% confidence level (c.l.) the hypothesis that a progenitor population with $\RGf\gtrsim 30\%$ has generated our data. A contribution of more
than  $40\%$ is ruled out at the $>95\%$ c.l. These results are largely
consistent with the analysis presented in \citet{2010sdss}, and with the
upper limits placed by~\citep{2010Natur.463..924G} in SNIa in elliptical
galaxies. Accounting for all angles, with only 
$\sim200$ data points in the 10 days that would be affected by the shocking
effect, we would expect $<40$ data points to be affected
significantly by the effect we are probing, even if \emph{all}
companions were RGs. This is therefore a small number statistics
problem, in the presence of noise in both data and models, and it is not surprising that we have a limited ability to
 place strict limits to the contribution of RG progenitors this way. However we will obtain stronger limits by including considerations on the color bias in the shocking excess in the next section (Section~\ref{sec:ksc}).

\begin{figure}[!h]
\centerline{\includegraphics[width=0.48\textwidth]{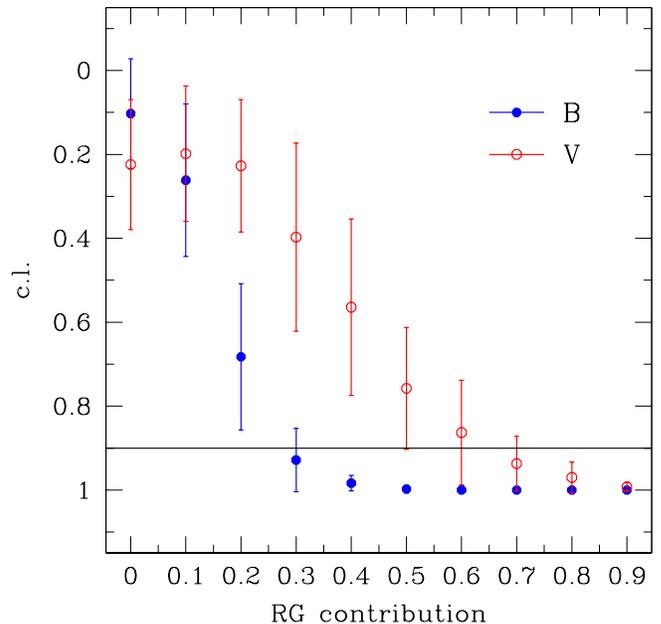}}
\caption{Results from 2-sample K-S tests applied to the early rise
 portion of the composite lightcurves from the SNLS. 
We plot the confidence level (c.l.) at
 which we can reject the null hypothesis that the true data comes from
 the same distribution of progenitor that generated the simulated data,
 as a function of $\mathrm{RG}_\mathrm{frac}$. The 90\% c.l. is marked with a solid black line. The true and synthetic population are increasingly less likely to come from the same distribution as the RG contribution to the mixture of 
 synthetic progenitors is increased: RGs are unlikely to be common progenitors of \SNIae. The
 effect is more pronounced -- and strictly monotonic -- in B band (blue) than in V band (red) because
 of the color bias in the shocking excess. A $\RGf<0.3$ is allowed
 within the $1\sigma$ error bar while larger contributions are ruled
 out at 90\% c.l. or greater}
\label{fig:ks}
\end{figure}

Noticeably, despite the noise, the probability that the true and synthetic distribution of data points come from the same progenitor distribution clearly decreases monotonically as we increase the contribution of RG systems, particularly in the \BRF band (Figure~\ref{fig:ks}). This strongly suggests a minimal contribution of RGs to \SNIa~progenitors. Similarly, an almost monotonic decrease in probability (increase in c.l.) is evident in the \VRF band, though less pronounced. This is expected, on account of a smaller signature of shocking in redder bands (see Figure~\ref{fig:models}).

\subsection{K-S chromatic test}\label{sec:ksc}
In this section we investigate the chromatic bias in the shocking footprint. In
absence of shocking, the expected time behavior of the SN explosion is a
parabola, similar in \VRF and \BRF band. Thus, in the rise portion of
the lightcurves we would expect points drawn from a set of SNe to come,
statistically speaking, from the same distribution in \VRF and \BRF
band. However, in the K10 simulations (see Section
\ref{sec:models}) the \VRF and \BRF time behavior differ dramatically in
the rise portion of the lightcurve in the presence of RG progenitors. We again perform a 2-sample K-S test. This time we want to asses the similarity of the B and V  populations of early-rise data, so for the SNLS data and  we compare the B and V channel with a K-S test, and we do the same for each synthetic population. 

We find that the hypothesis that rest-frame \BRF and \VRF populations of data points from the
composite true SNLS lightcurves, day 0-10 after explosion, come
statistically from the same distribution  can only be rejected to $<5\%$
c.l., or equivalently that the hypothesis that the two channels come from the same distribution has a $\pv \sim~0.95$.

We compare the synthetic B and V lightcurves, and find, as expected, that the K-S number increases with the increasing RG contribution: the probability that the B and V synthetic data come from the same distribution decreases as more RG progenitors are used in the progenitor mix.

\begin{figure}
\centerline{\includegraphics[width=0.48\textwidth]{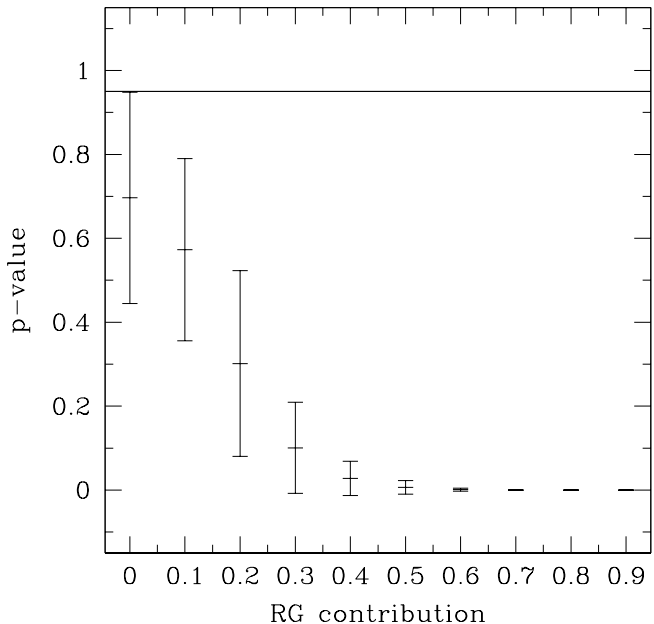}}
\caption{Results from 2-sample K-S tests performed to compare the early rise portion of the V and B band lightcurves. 
The solid line represents the \pv~of the hypothesis that the true
 \VRF and \BRF lightcurves come from the same distribution: $\pv \sim
 0.95$; below the line the B and V would diverge more than the SNLS rest-frame B and V do. The
 points represent the \pv's for our synthetic B and V data, as a function of the RG contribution in the progenitor
 mixture. That is: \pv~of the hypothesis that, for each RG contribution, the
 B and V band synthetic data came from the same distribution (where the
 \pv~is 1 minus the rejection level of the hypothesis). The $1\sigma$ error bars plotted are obtained generating 100  synthetic populations and accounting for the errors in the SNLS data and in the K10 simulations. 
Only the synthetic population that contains no RG progenitors is
 consistent at the $1\sigma$ level with the data. A $\RGf>10\%$ is ruled out
 at the $2\sigma$, and a $\RGf>20\%$ at the $3\sigma$ level.}
\label{fig:kscolor}
\end{figure}

The results of this test are plotted in Figure~\ref{fig:kscolor}. Within
$1 \sigma$ error bars, only the population with \emph{no
contribution from RG progenitors} is consistent with the SNLS data.  
We rule out a contribution $\RGf \gtrsim 10\%$ at $\sim2 \sigma$, and
greater than 20\% at $>3\sigma$ level.
\subsection{Color distributions}\label{sec:color}
\begin{figure}[!ht]

\centerline{\includegraphics[width=0.35\textwidth]{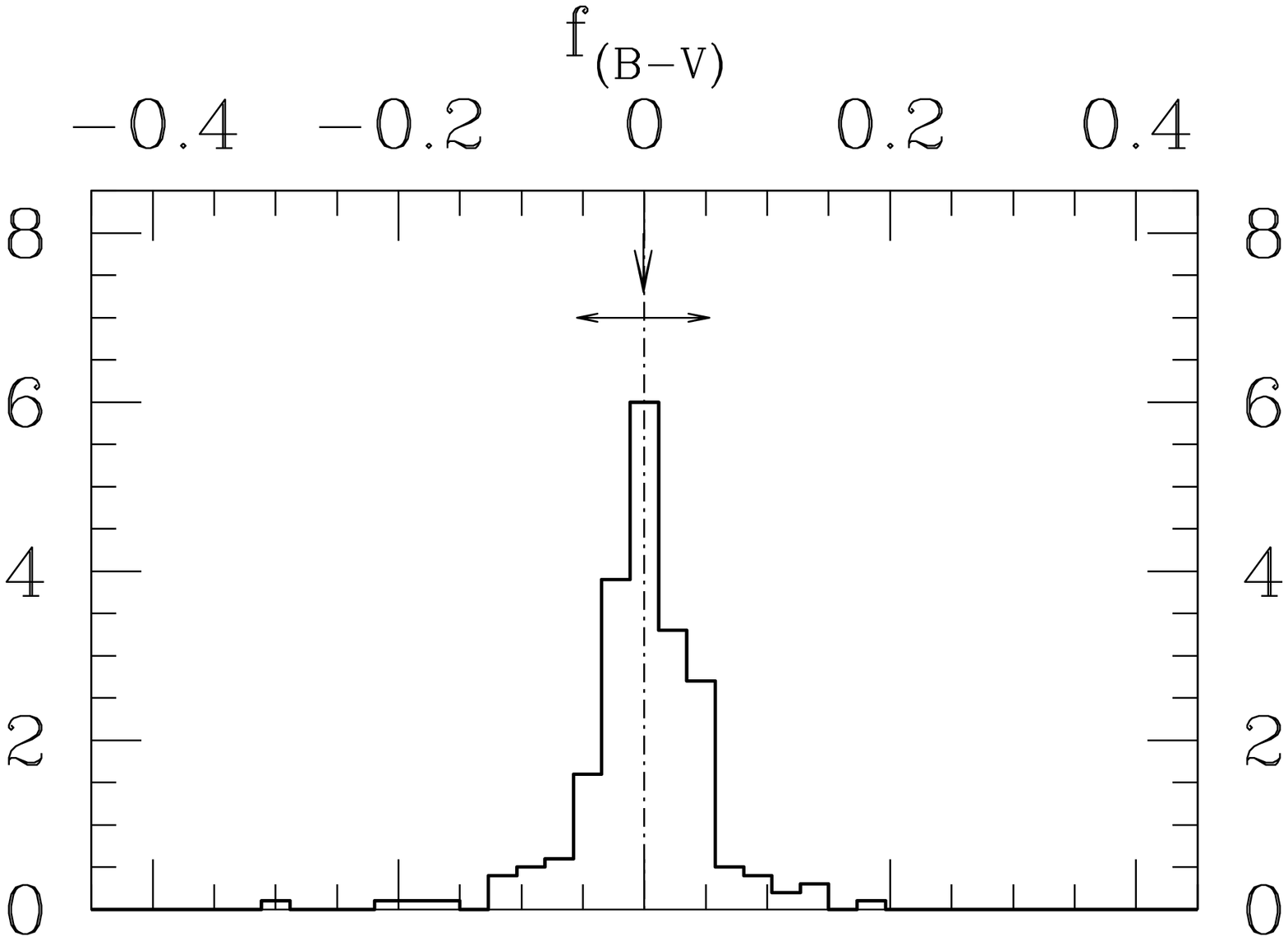}}
\centerline{\includegraphics[width=0.35\textwidth]{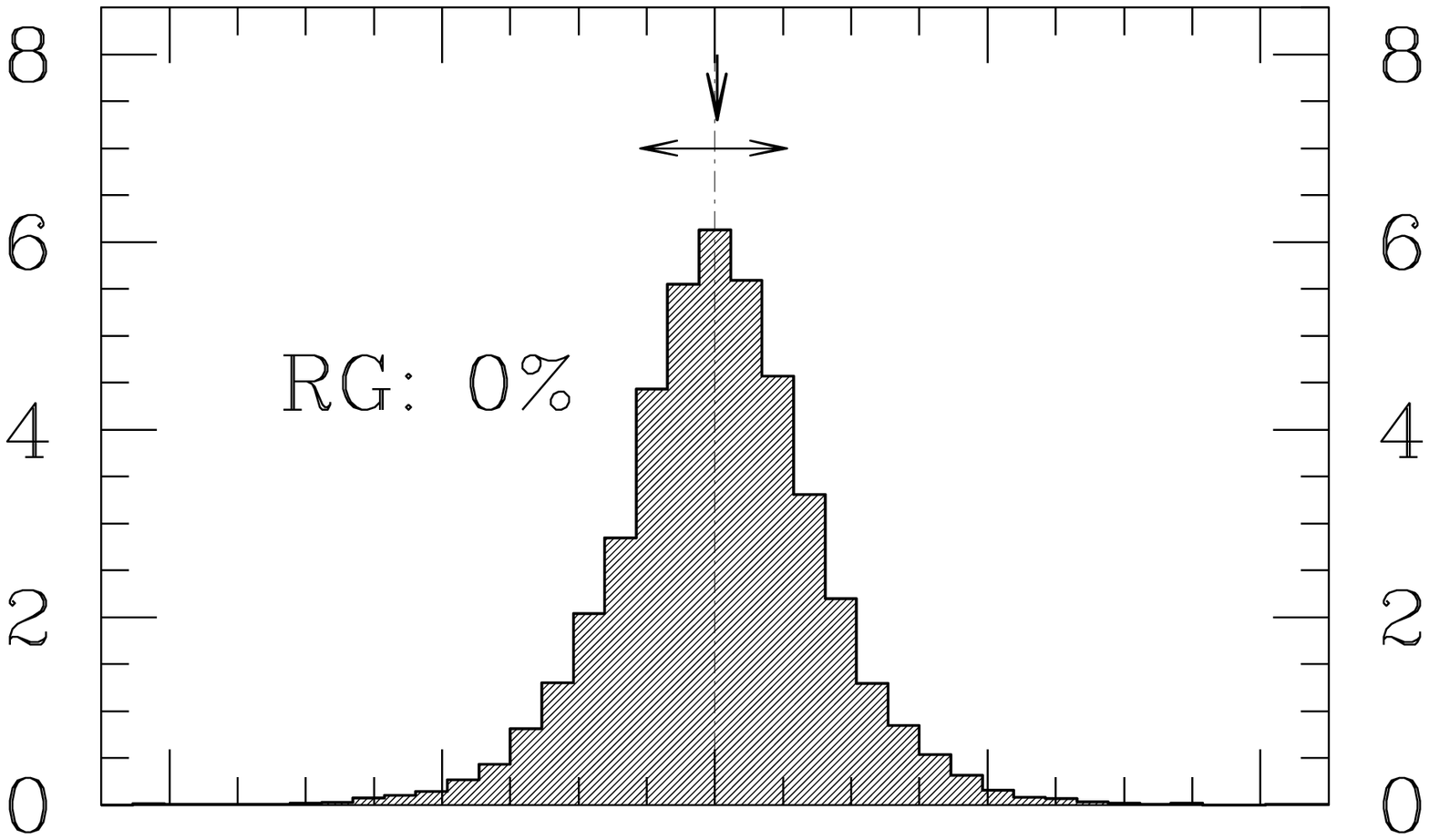}}
\centerline{\includegraphics[width=0.35\textwidth]{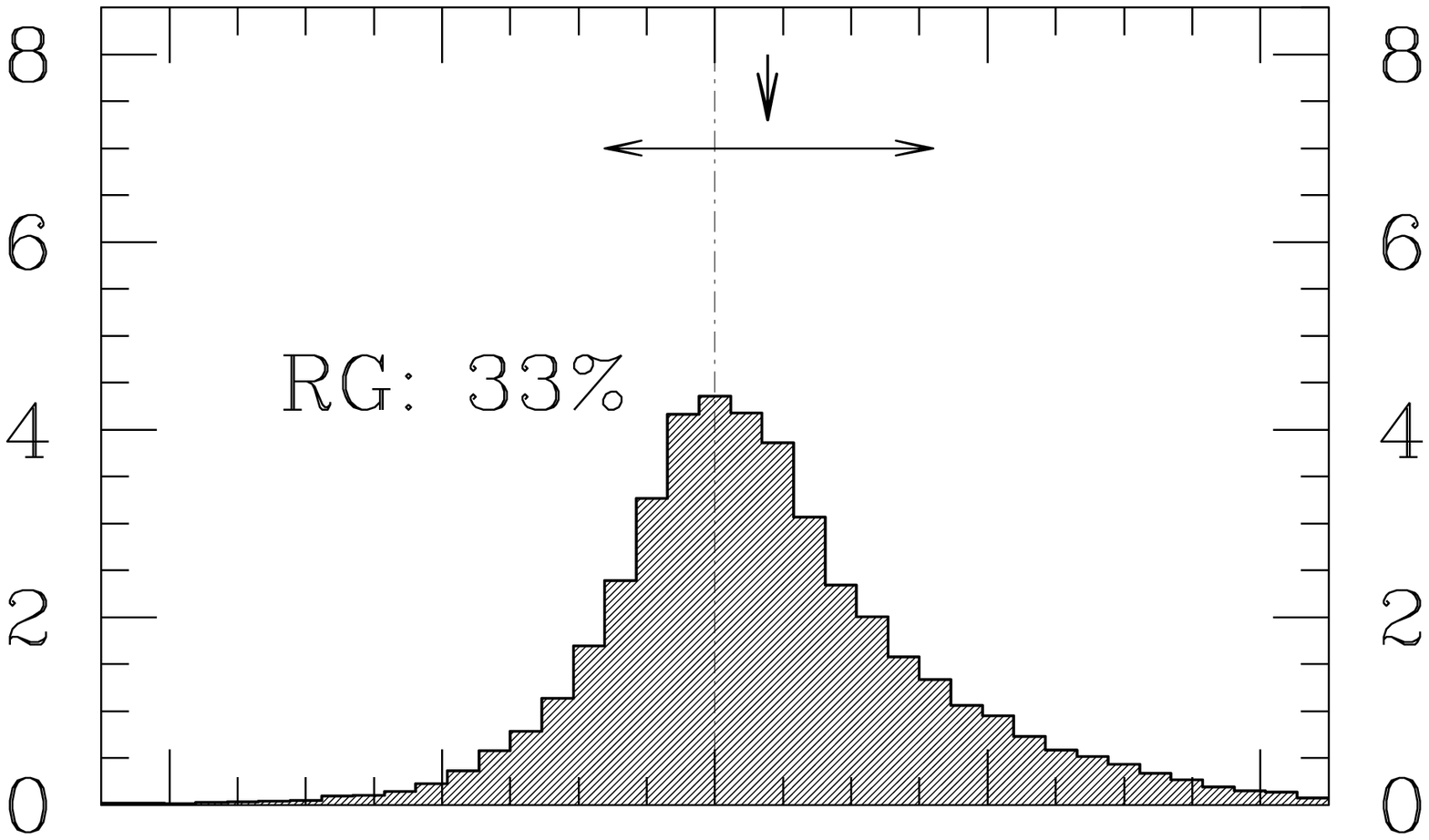}}
\centerline{\includegraphics[width=0.35\textwidth]{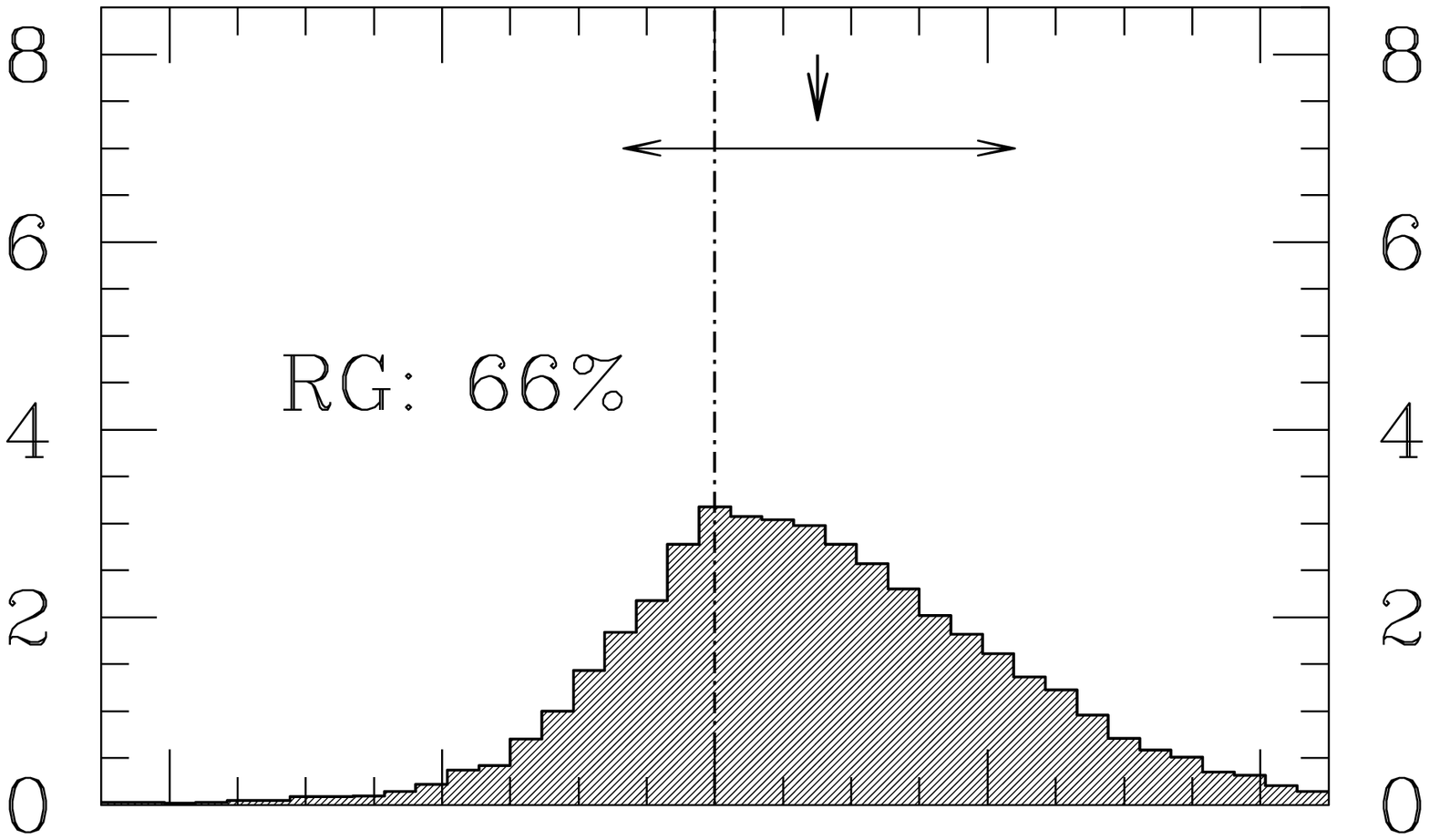}}
\centerline{\includegraphics[width=0.35\textwidth]{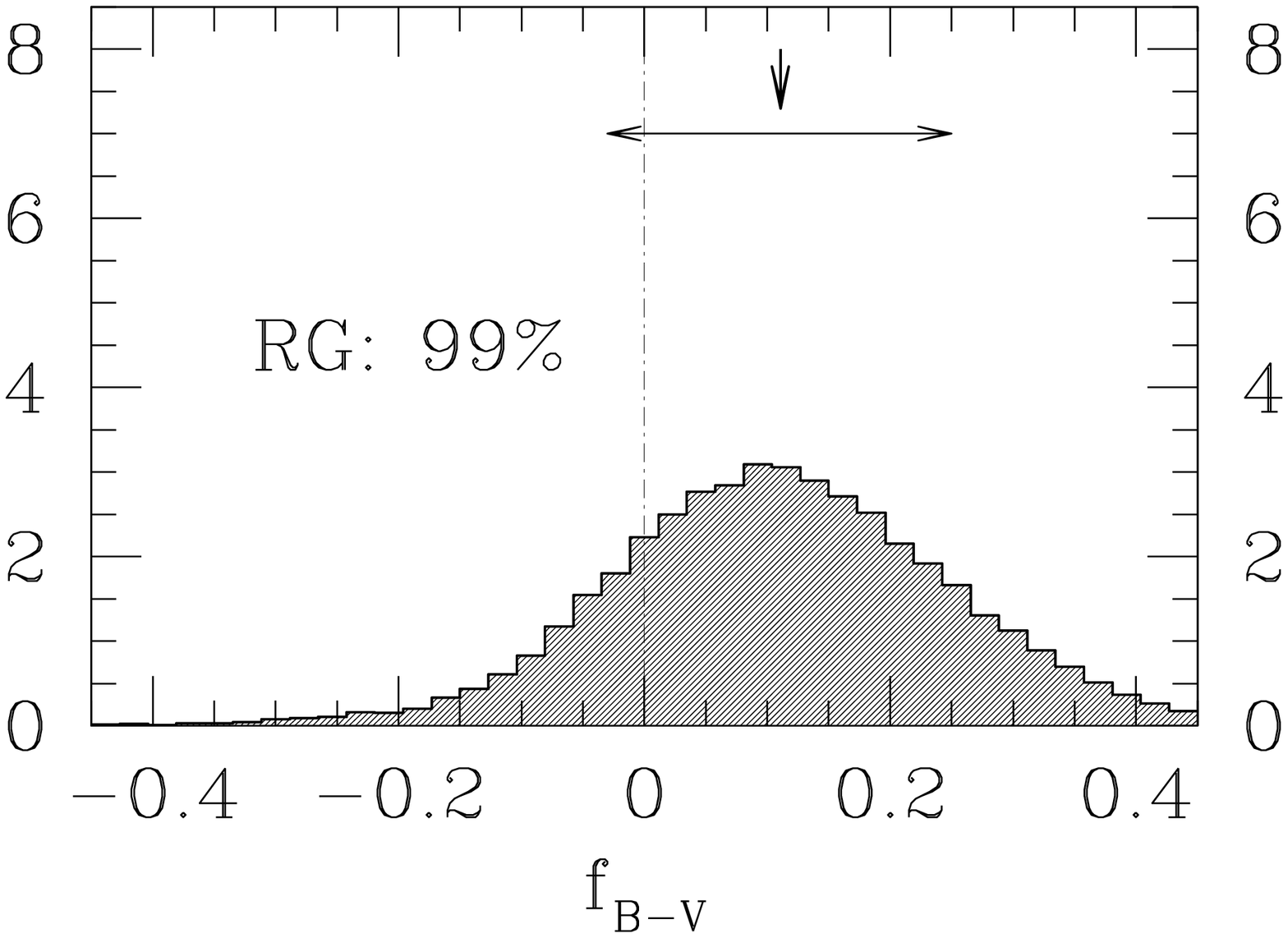}}
\caption{\BmV distribution for true (top) and simulated data (four bottom plots). 
The synthetic populations are generated from progenitor mixtures with, from the second to the last plot, \RGf=0, \RGf=33\%, \RGf=66\%, and \RGf=99\%. The mean for each distribution is plotted as a vertical arrow, and the standard deviation of each population as a double-ended arrow above the distribution. A dotted-dashed line shows the location of \BmV=0. B and V are normalized at peak, thus no color excess is expected in absence of shocking. Each synthetic population is a factor 100 larger than the true population, to minimize the Poisson noise.}
\label{fig:BmV}
\end{figure}
In standardizing our lightcurves we have chosen \emph{one} stretch value for each lightcurve to be applied to all rest-frame bands. We ask the question: could there be a correlation between \VRF and \BRF that would interfere with the result of our K-S chromatic test (see Section~\ref{sec:ksc}). Suppose some data points in the region $\trf<-10$~days, which is not fit to the template,
have more (or less) flux than the template, so that if we included
those points when fitting the lightcurve to the template we  would have generated different fit parameters, and a
different value for the stretch or day of maximum; in this case both the \BRF and \VRF
lightcurves would show flux in excess (deficit) of the template in the
shock region. We might then see a correlation in our \BRF and \VRF
composite lightcurves.
 Any such correlation would systematically lower the K-S number (higher the
\pv~for the null hypothesis) found for the true data.
Meanwhile our simulated lightcurves are generated independently in B and
V bands.  Since in our chromatic K-S test even the simulations which use
no RGs among the progenitors are consistent with our true data at the
$1\sigma$ level, but show a lower \pv, indicating a greater discrepancy
between B and V band, we further explore this possibility, seeking an independent confirmation of what we see in the K-S tests beyond this possible systematic correlation. 

We test the color of our true and simulated lightcurves by taking $\BmV$
to be the
difference of the B and V flux after normalizing each channel at
peak. Under the assumption that in absence of shocking the rise portion
of the lightcurve would follow a parabolic behavior identical in both
bands, diverging only at $\trf\geq-9$ as modeled in the {\tt Conley09f} 
template, the distribution of \BmV values should be consistent with 0 for our composite lightcurves, while the effect of shocking would produce a distribution of \BmV with a positive mean, and a large standard deviation.

For every flux point in each normalized, standardized rest-frame \BRF lightcurve, we subtract the flux of the closest rest-frame \VRF data point, within $\Delta \trf < 0.2~\mathrm{days}$. Similarly,  we generated synthetic colors for different RG fractions, by creating pairs of V and B synthetic lightcurves, accounting as usual for the typical error bars in the data and in the models. We derive the distribution of colors for both true and simulated data. 

The \BmV distributions are plotted in Figure~\ref{fig:BmV}. The top panel shows the \BmV distribution for true data.
There is no blue excess in flux in the true color: in fact the distribution has a  mean of $\mu~\sim ~-8 \times 10^{-4}$, a median $\sim 0.002$, and a standard deviation $\sigma \sim 0.054$: statistically consistent with a random distribution around 0. 

The distributions generated from simulated lightcurves are shown below
the distribution for true data in Figure~\ref{fig:BmV}, for RG contributions 
$\RGf~=~ 0\%,~33\%,~ 66\%,~ \mathrm{and}~ 99\%$, plotted from the top to
the bottom. Each distribution is generated from a factor of 100 more points than the true color distribution and is thus minimally noisy. The mean of the distribution increases as we increase \RGf and the distributions get increasingly asymmetric, weighted toward positive values of \BmV (bluer color).
The synthetic distributions generated with no RGs (\RGf=0\%) has moments that are extremely similar to those of the true color distribution: $\mu~\sim~2\times 10^{-3}$,  median $\sim 0.001$ and   $\sigma \sim 0.077$. Once again, this shows that the distribution of colors in the SNLS data is compatible with minimal -- or no -- contribution of RG to \SNIa~progenitors, confirming the results obtained from the K-S tests.      

\begin{figure*}[!t]
\centerline{\includegraphics[width=0.48\textwidth]{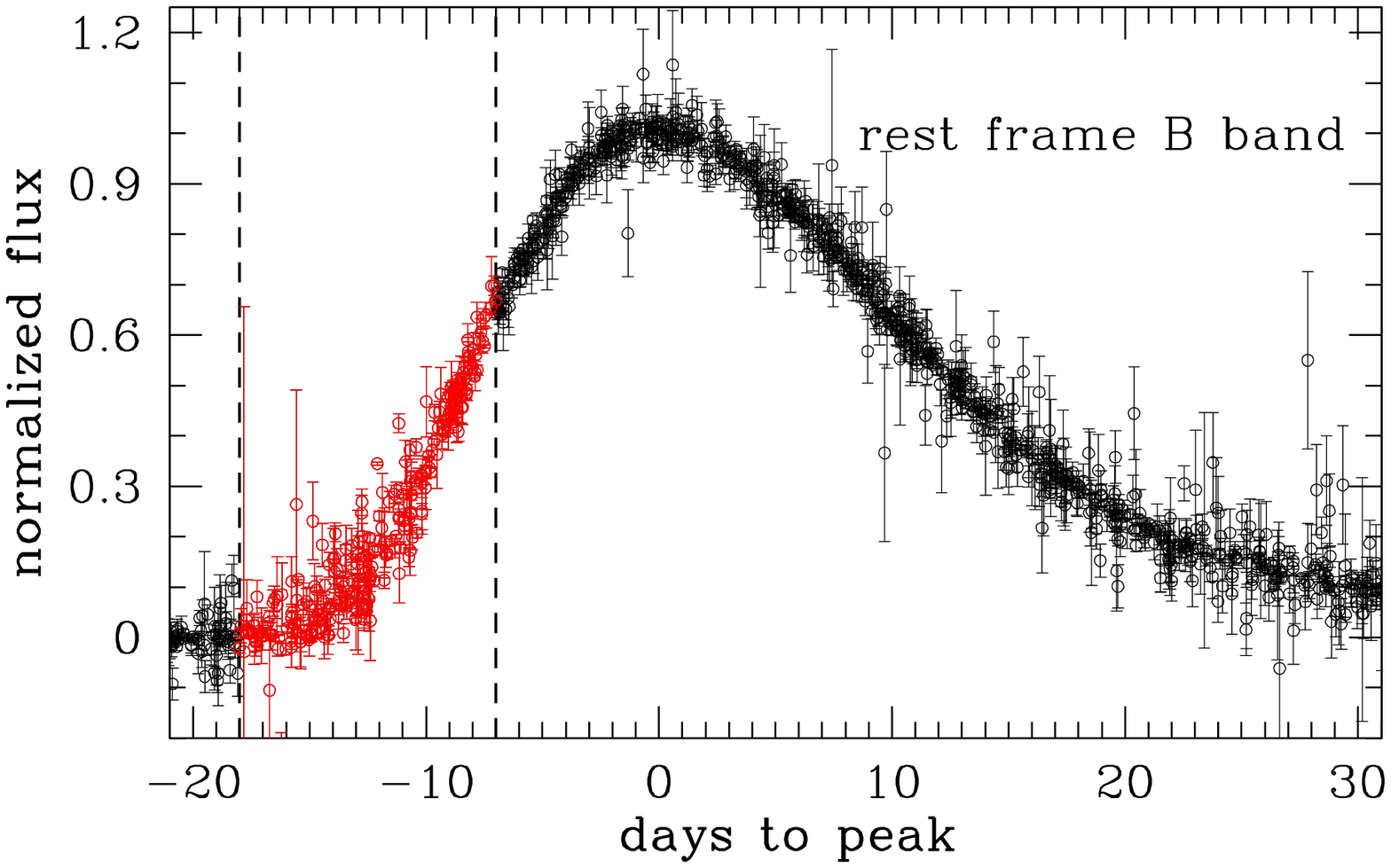}\includegraphics[width=0.48\textwidth]{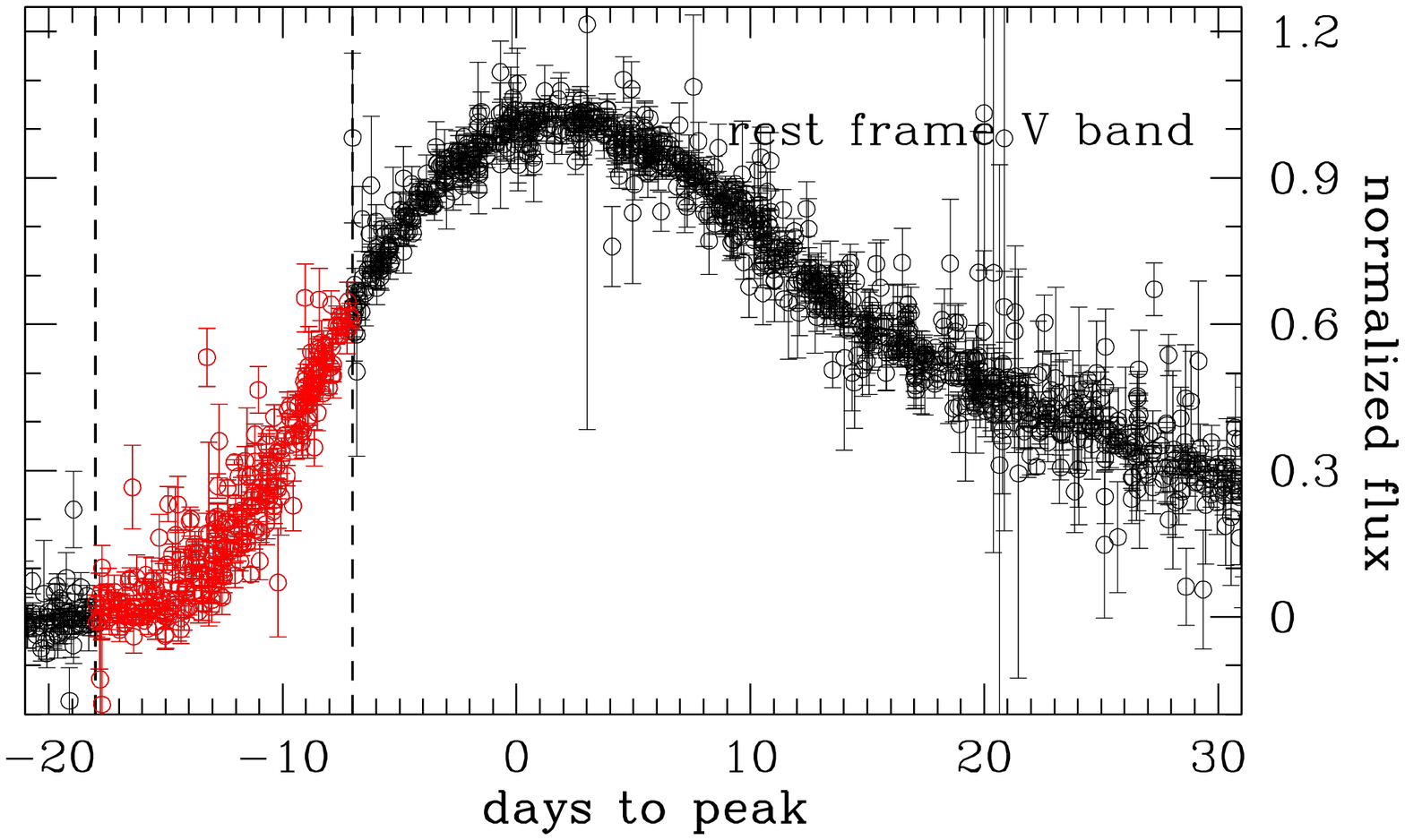}}
\caption{Composite B (left) and V (right) SNLS lightcurves containing both confirmed and unconfirmed \SNIae. Points plotted in red, and enclosed by vertical lines, are those which may be affected by the excess induced by the presence of a companion. }
\label{fig:complc_uncf}
\end{figure*}

\section{Photometrically selected \SNIae}\label{sec:unconf}
The excess due to shocking of the SN ejecta affects the early time
domain photometric and spectral behavior of the \SNIa~explosions. Since
in surveys such as SNLS and SDSS \SNIa~are identified by their early
lightcurves, and thus an explosion is followed up spectroscopically only
if it is thought to be a SN explosion, an interesting question is
whether this early effect might have lead to the rejection of phenomena
that indeed were \SNIa, but deviated from the expected early behavior
on account of shocking. In~\citet{2010sdss}, a subset of unconfirmed
\SNIa~is visually inspected and no such effect is found. We investigate
905 SNLS lightcurves with some redshift information, either
spectroscopic or photometric.  We exclude likely or known AGN, variable
stars, and core-collapse (CC) SNe.  In order to avoid contamination from
unidentified SNe II, Ib, or Ic, we also apply cuts in stretch and color
space.  In particular, CC SNe show a different average color
than SNe Ia, and color constraints eliminate them from the sample.  A
detailed discussion of photometric selection of \SNIae~in the SNLS data
can be found in \citet{Bazin10}. 
We thus believe our new dataset has minimal
contamination from non \SNIa~events.  Our new dataset contains 336
lightcurves before our cuts are applied (see
Section~\ref{sec:data}), and 110 after. Our new composite lightcurves
contain 251 points in rest-frame \BRF and 270 in rest-frame \VRF in the
region of interest: \trf=-17.4 to -7.4 days to peak
(Figure~\ref{fig:complc_uncf}).

\begin{figure}
\centerline{\includegraphics[width=0.48\textwidth]{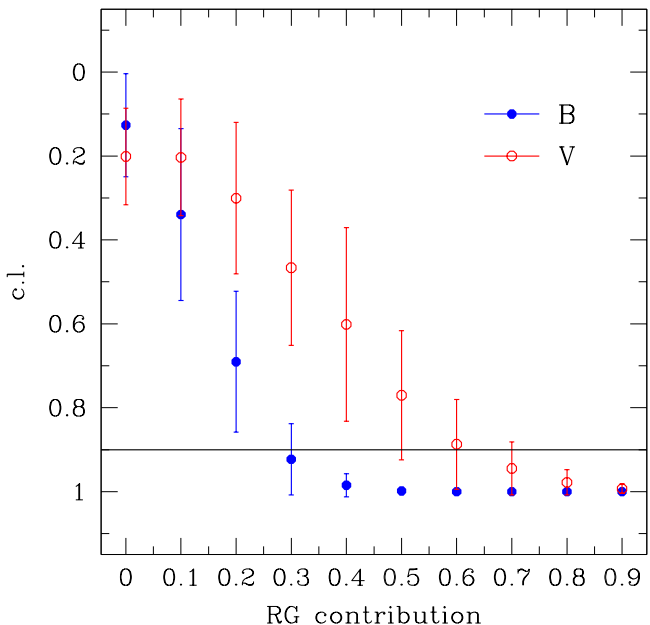}}
\caption{Results from 2-dimensional K-S applied to the early rise portion of the composite lightcurves. The Figure reproduces Figure ~\ref{fig:ks} for the extended \SNIa~set, not limited to spectroscopically confirmed events.}
\label{fig:ks_unconf}
\end{figure}

\begin{figure}
\centerline{\includegraphics[width=0.48\textwidth]{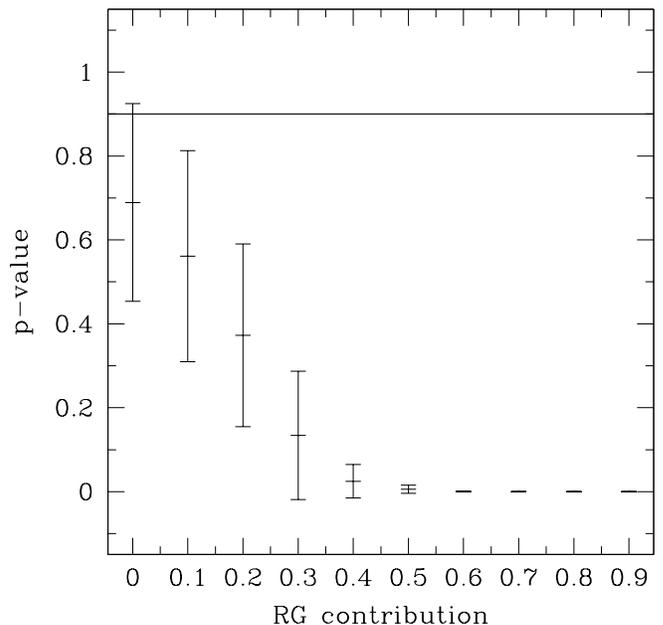}}
\caption{Results from 2-dimensional K-S applied to the early rise
 portion of the  color composite lightcurves. This Figure reproduces Figure~\ref{fig:kscolor} for the extended \SNIa~set, not limited to spectroscopically confirmed events.}
\label{fig:ks_color_unconf}
\end{figure}

We repeat the K-S tests applied earlier to the extended \SNIa~set and
find that
the statistics confirm the upper limits set to the contribution of RG binary systems to \SNIa~explosions (Figures~\ref{fig:ks_unconf} and ~\ref{fig:ks_color_unconf}). The K-S test of the composite lightcurve in each B and V with the respective synthetic lightcurves is entirely consistent with the test for the spectroscopically confirmed \SNIa~subset, and consistent with minimal or no contribution of RG to the \SNIa~progenitors.

\section{U band data}\label{sec:uband}
\begin{figure}
\centerline{\includegraphics[width=0.48\textwidth]{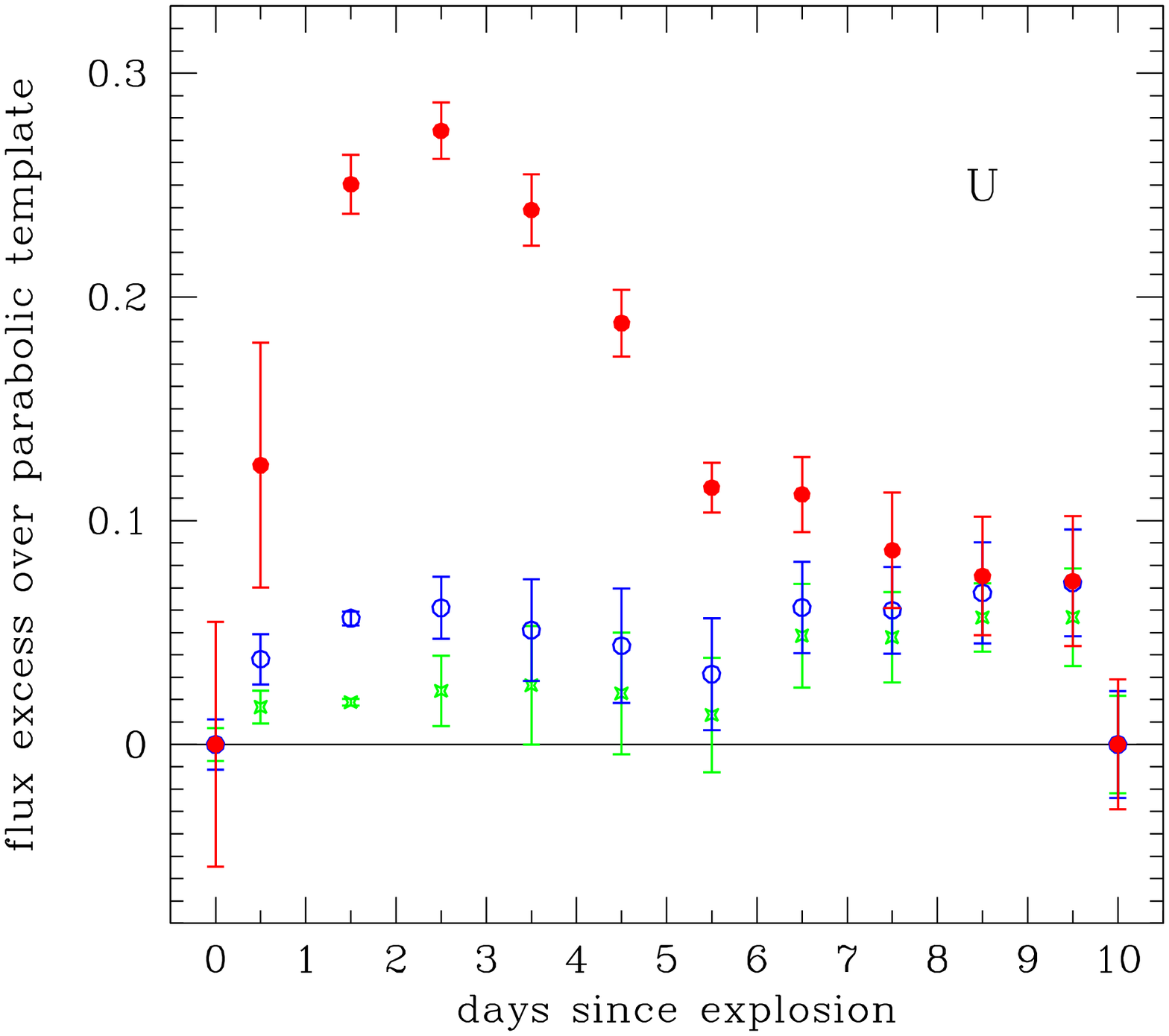}}
\caption{Excess over a template caused by shocking in U rest-frame band, according the the theoretical models presented in K10: the figure reproduces
 Figure~\ref{fig:excessBV} for the \URF band. The angle averaged excess,
 in units of peak luminosity, is shown for a RG with red 
 filled circles, for a $6~M_\odot$ with blue empty circles and for a
 $2~M_\odot$ with green crosses.}.
\label{fig:excessU}\end{figure}

As described in Section~\ref{sec:models}, the excess due to
shocking is more prominent at bluer wavelengths. In \URF band, the models by
K10 predict an excess over
a nominal template  more pronounced by roughly 20\% over the B band for the RG case, averaged over all angles and over the first 10 days after explosion. The
prediction for the angle averaged excess in \URF band is shown in
Figure~\ref{fig:excessU}. We thus extend our analysis to the rest-frame
\URF band data, to see if a stronger constraint can be placed to the
contribution of RG to the \SNIa~ progenitor population. 
Three years of SNLS spectroscopically confirmed lightcurves are processed
as described in Section~\ref{sec:data}, in order to generate rest-frame
\URF band lightcurves. The lightcurves are then selected if they pass similar
cuts to those described earlier:
\begin{itemize}
\item spectroscopically confirmed Type Ia SNe at redshift $z < 0.7$ (135 lightcurves)
\item the total reduced \chisq~for the SiFTO template fit, applied to
      epochs  $\trf > -10$ days, is better than 3.0 (130 lightcurves)
\item the error in the determination of the peak date is $\Delta d_\mathrm{max}<0.7$ days (117 lightcurves)
\item have at least three data points in rest-frame B, three data
      points in rest-frame V, and three in rest-frame \URF band  in the rise portion of the lightcurve, $-10 \leq \trf \leq0$~days, to ensure the quality of the pre-peak fit (57 lightcurves).
\end{itemize}

\begin{figure}
\centerline{\includegraphics[width=0.48\textwidth]{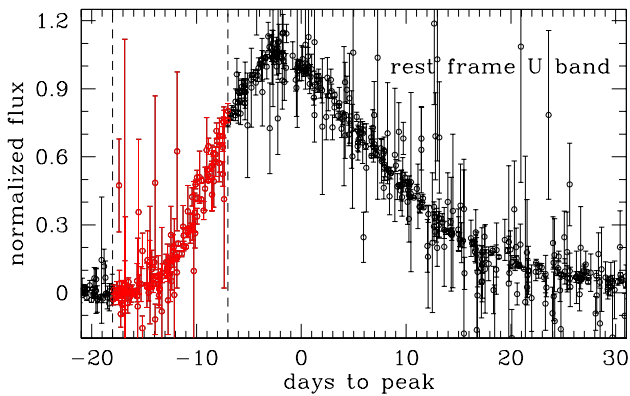}}
\caption{Composite U lightcurve from three years of SNLS data. Symbols
 are as described in Figure~\ref{fig:complc}.}
\label{fig:complcU}\end{figure}

The rest-frame \URF band composite lightcurve thus generated is plotted
in Figure~\ref{fig:complcU}, and it contains 662 points between days -20
and 40 from explosion, and 123 in the region of interest: $-17.4 \leq
\trf \leq -7.4$. 
Applying the latter cut, which is more restrictive
than the corresponding cut in our primary analysis, the
new composite \BRF and \VRF lightcurves contain, respectively, 152 and 161
data points in the region  $-17.4 \leq \trf \leq -7.4$. 
It is immediately evident that the \URF band composite lightcurve is
significantly noisier than the \BRF and \VRF composites, and it
contains 35\% fewer lightcurves, and roughly 40\% fewer relevant data
points.

\begin{figure}
\centerline{\includegraphics[width=0.48\textwidth]{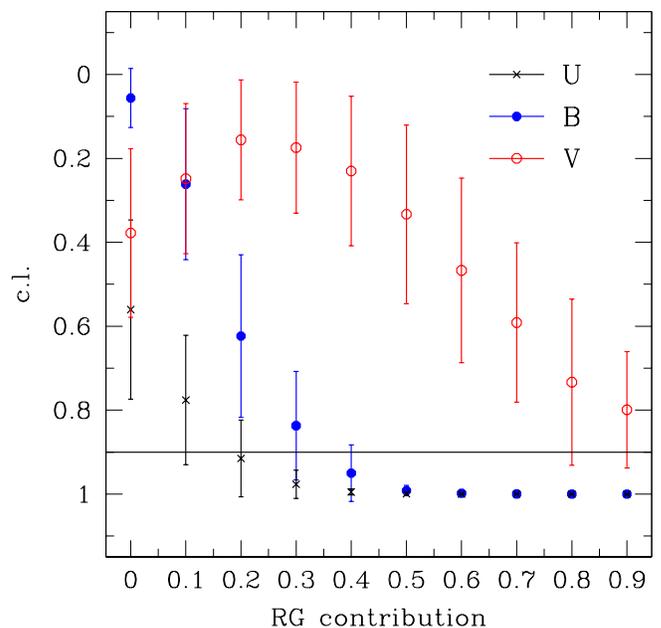}}
\caption{Results from 2-dimensional K-S applied to the early rise
 portion of the composite \URF (black crosses), \BRF (blue full
 circles), and \VRF (red empty circles) lightcurves; $1\sigma$ error
 bars are shown. This Figure
 reproduces Figure ~\ref{fig:ks} in the three color bands, and for the
 subset of lightcurves selected by the cuts described in Section~\ref{sec:uband}.
In U band, a RG fraction of 30\% and its $3\sigma$ error bars lie below
 the solid line, which indicates the 0.95 c.l of rejection of the hypothesis that the simulated and true data come from the same distribution. 
}\label{fig:ks_u}
\end{figure}

We generate synthetic \URF band lightcurves as described in
Section~\ref{sec:sim} and we reproduce the 2-sample, monochromatic K-S test we described in
Section~\ref{sec:onebandks} for the \URF band data. We find that the
true data distribution once again grows dissimilar from the simulated
distribution as more RG progenitors are included in the simulation
(Figure~\ref{fig:ks_u}). Using the \URF data a progenitor fraction
$\RGf>=30\%$ is ruled out at the $3\sigma$ level.
The results of the single band K-S test for the \BRF and \VRF for the subsample of
lightcurves that pass the new set of cuts are also plotted, and they are
entirely consistent, though with larger errors on account of the smaller dataset size, with the results obtained in
Section~\ref{sec:onebandks}. 

Note that the \URF band data, even in absence of RG in the simulated
data, appears from a K-S test to be different at the $2\sigma$ level
from the parabolic-rise model (i.e. the 2$\sigma$ limit of the c.l. of
rejection of the null hypothesis that synthetic and true data come from
the same distribution is below c.l. = 0 for all synthesized
populations).  The simulated lightcurves are generated as described in
Section~\ref{sec:sim}, and an adiabatic (parabolic) expansion is
postulated up to 6 days after explosion. However, effects of line
opacity and dispersion in the spectra at wavelengths bluer of $\lambda=
400 ~\mathrm{nm}$, as described in \citet{Ellis08}, affect the \URF
lightcurve, and may modify it from the our simple model
prediction. This can explain the
relative low correlation between our true and simulated data, which is
revealed by the K-S test. For this reason we are reluctant to apply the
chromatic K-S test describe in \ref{sec:color} to the U band data, as
our assumption that the U and B, or U and V lightcurves would have
identical early rise behavior might not hold here.  As
a matter of visualization though, we reproduce Figure~\ref{fig:BmV} for
the \UmV and \UmB data (see Section~\ref{sec:color}). As in all other
cases considered, the distribution of simulated color gets increasingly
different from the distribution of true data as we increase the
contribution of RG in the progenitor mix for our simulations.
\begin{figure*}

\centerline{\includegraphics[width=0.35\textwidth]{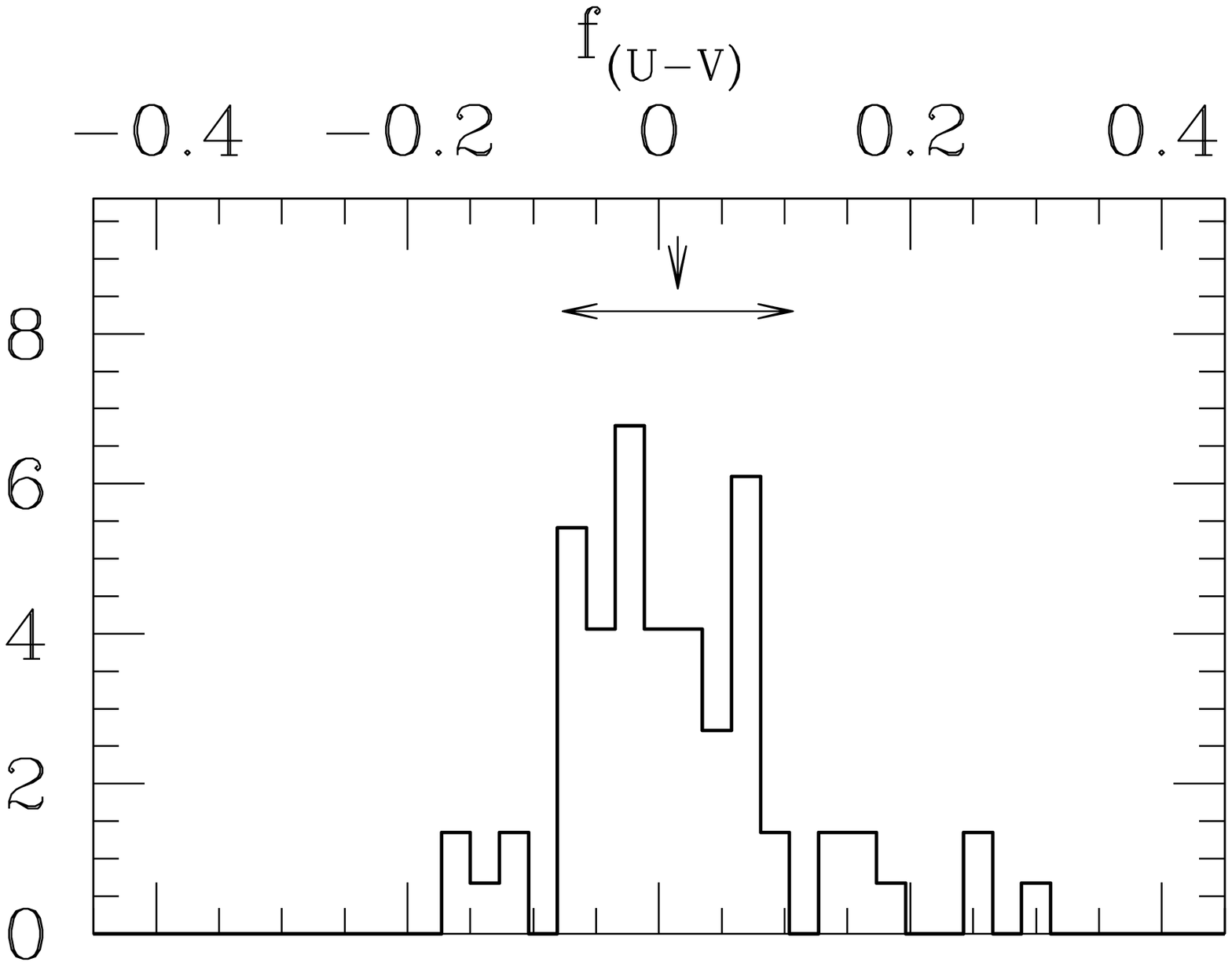}\includegraphics[width=0.35\textwidth]{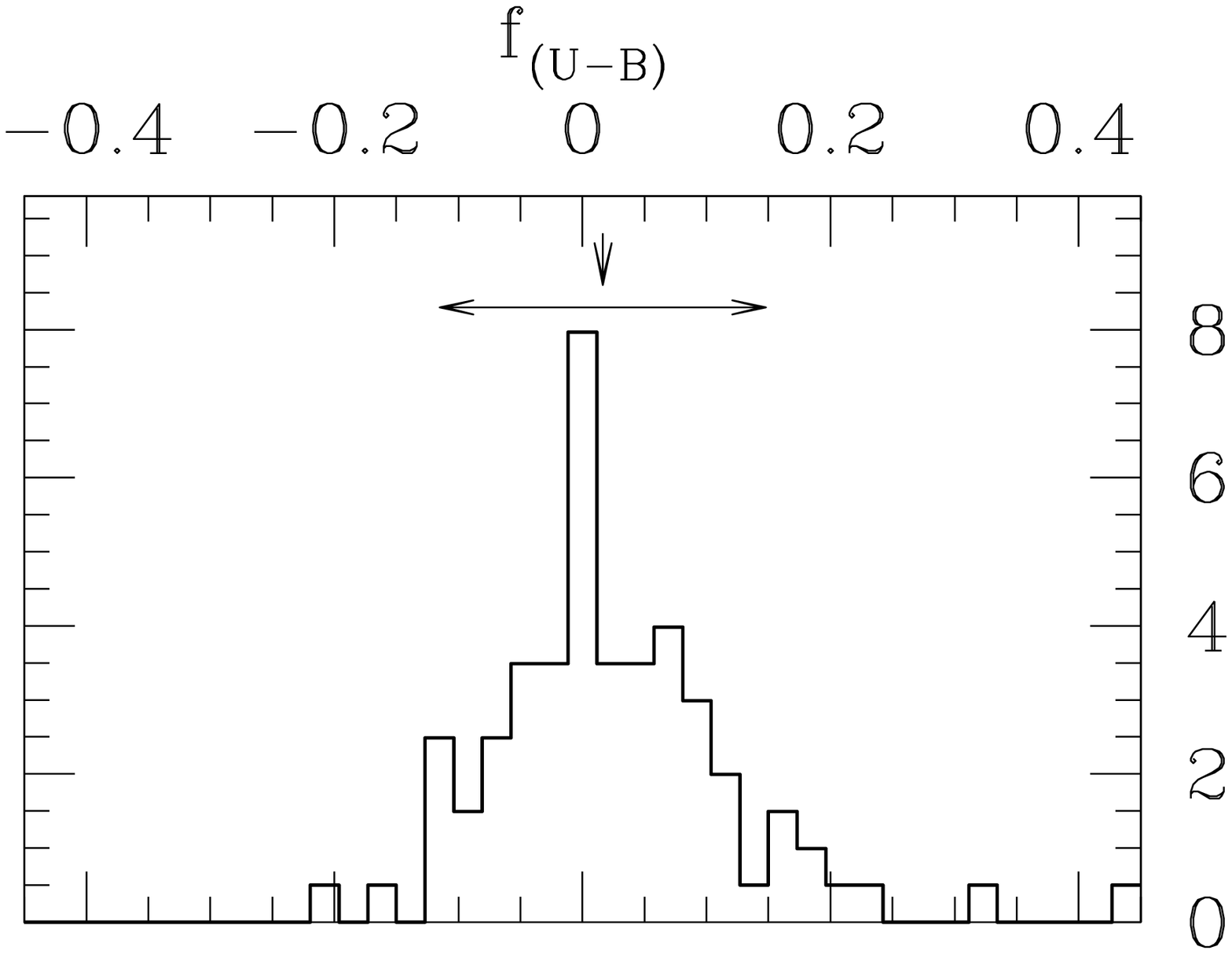}}
\centerline{\includegraphics[width=0.35\textwidth]{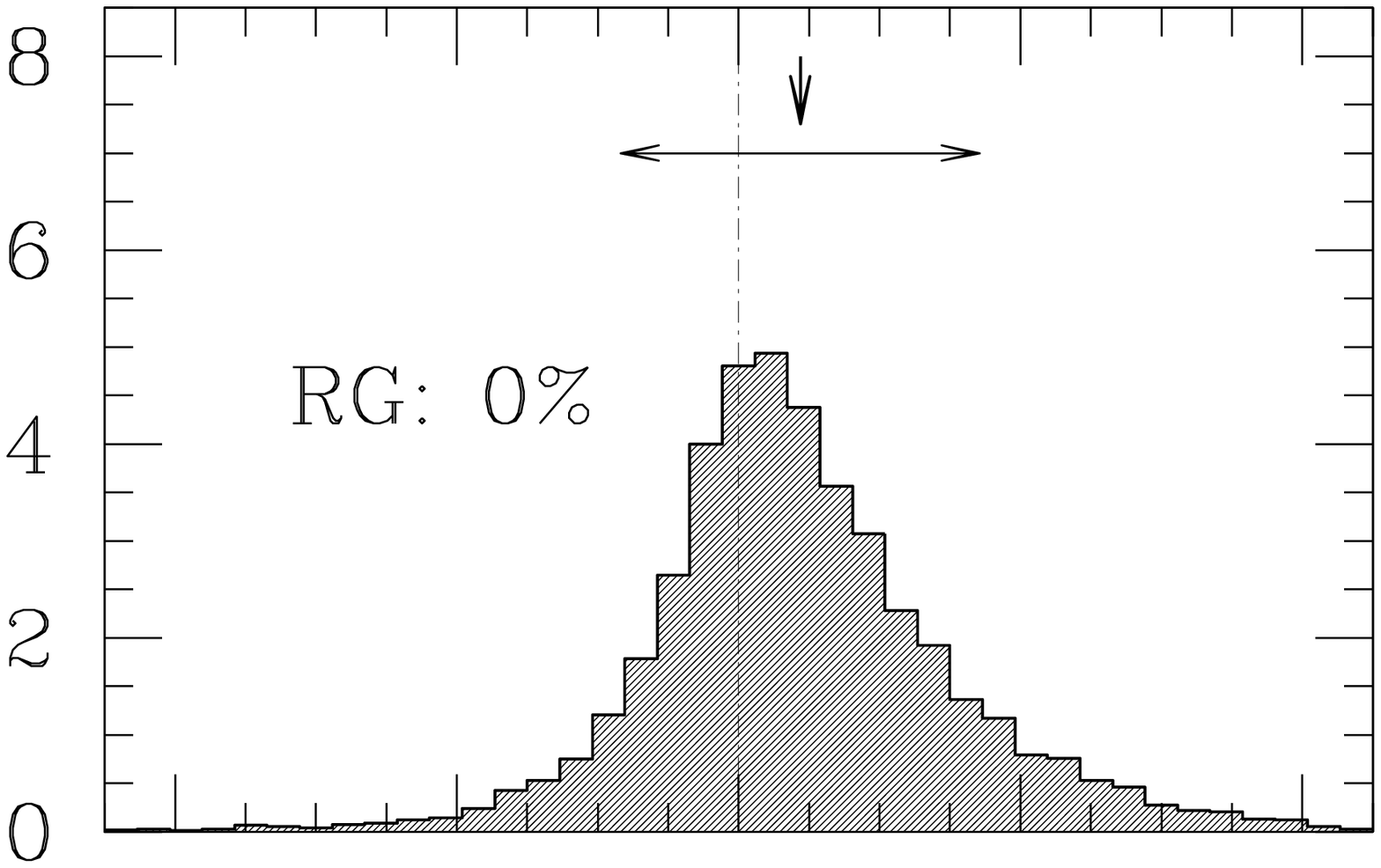}\includegraphics[width=0.35\textwidth]{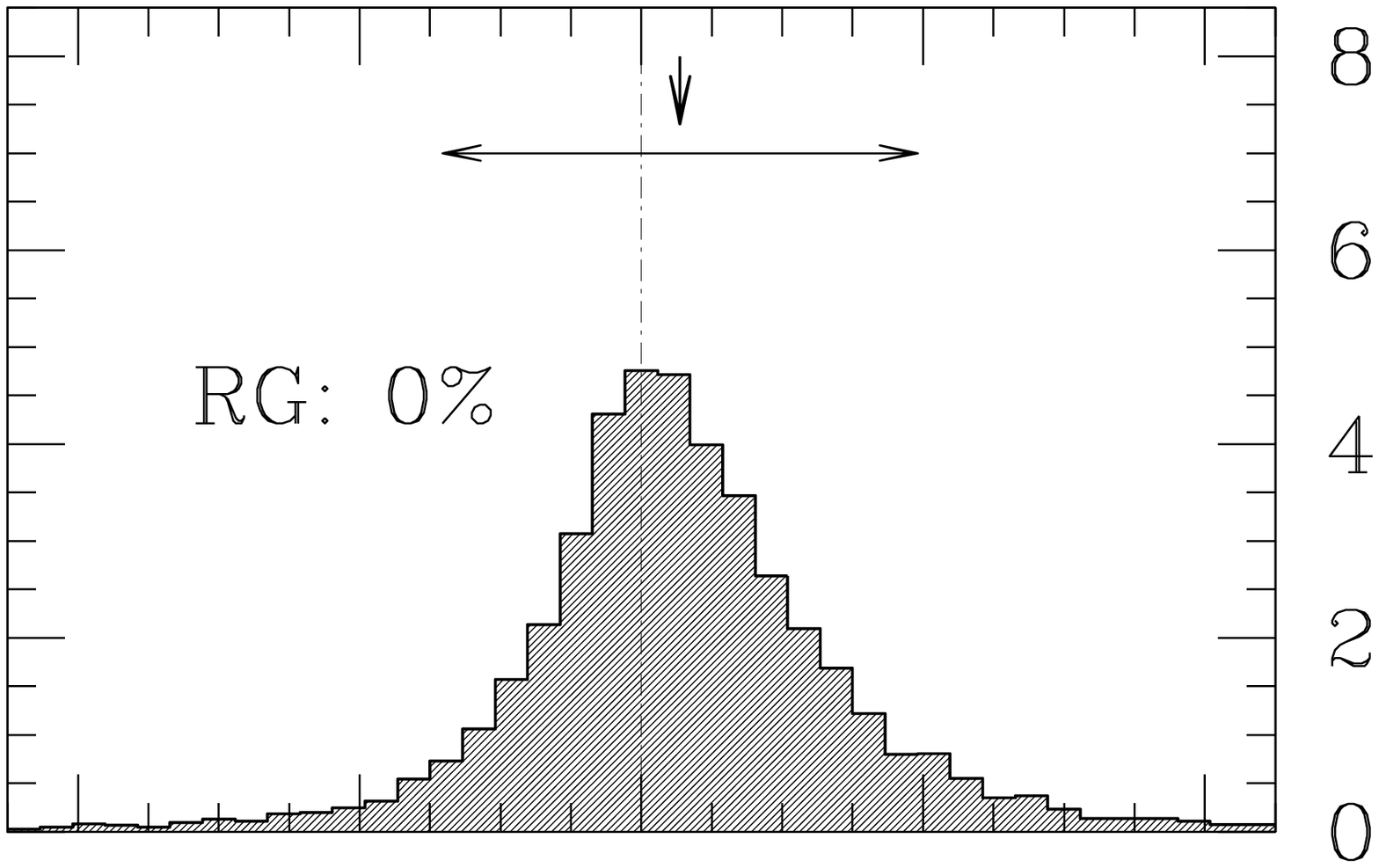}}
\centerline{\includegraphics[width=0.35\textwidth]{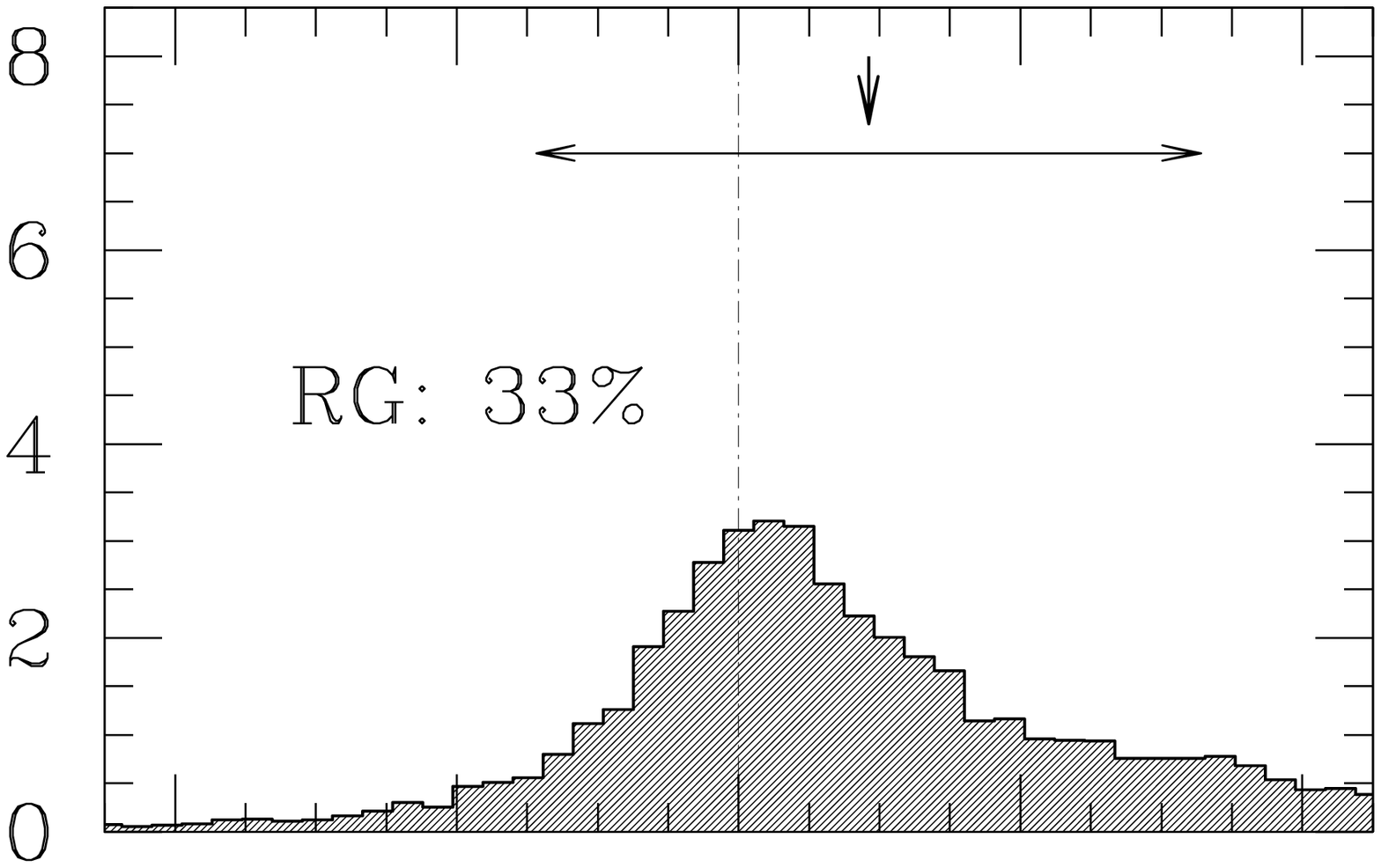}\includegraphics[width=0.35\textwidth]{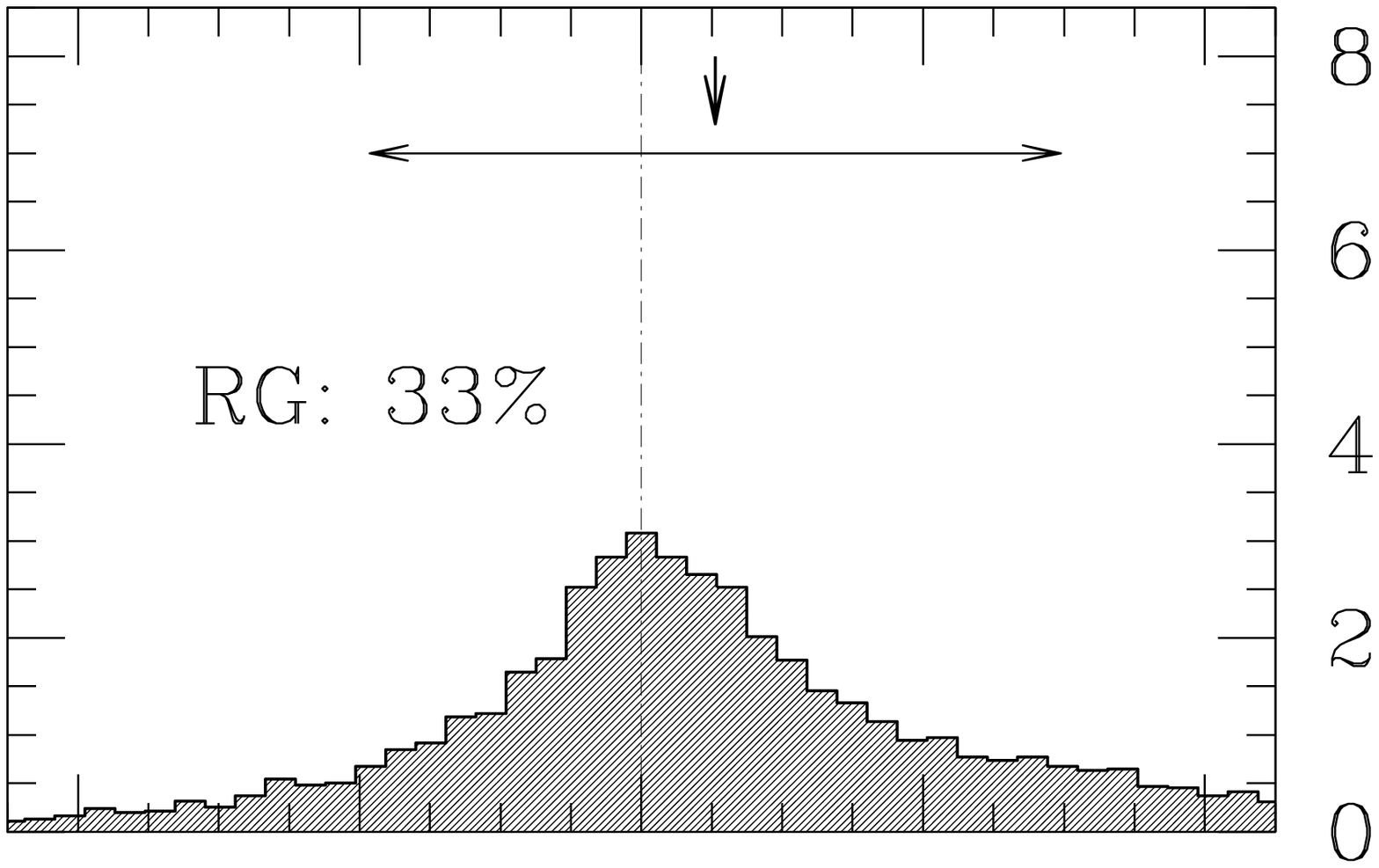}}
\centerline{\includegraphics[width=0.35\textwidth]{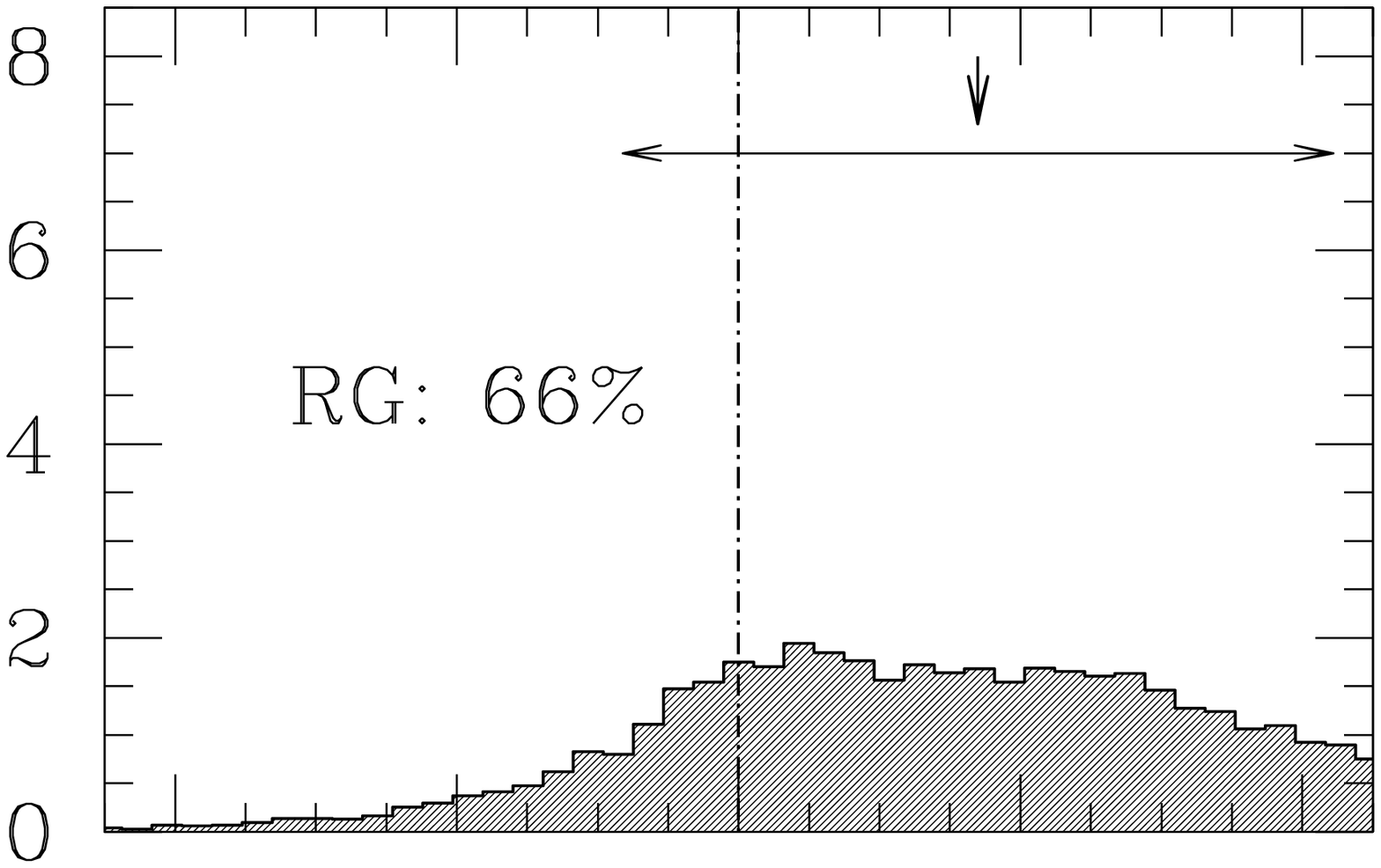}\includegraphics[width=0.35\textwidth]{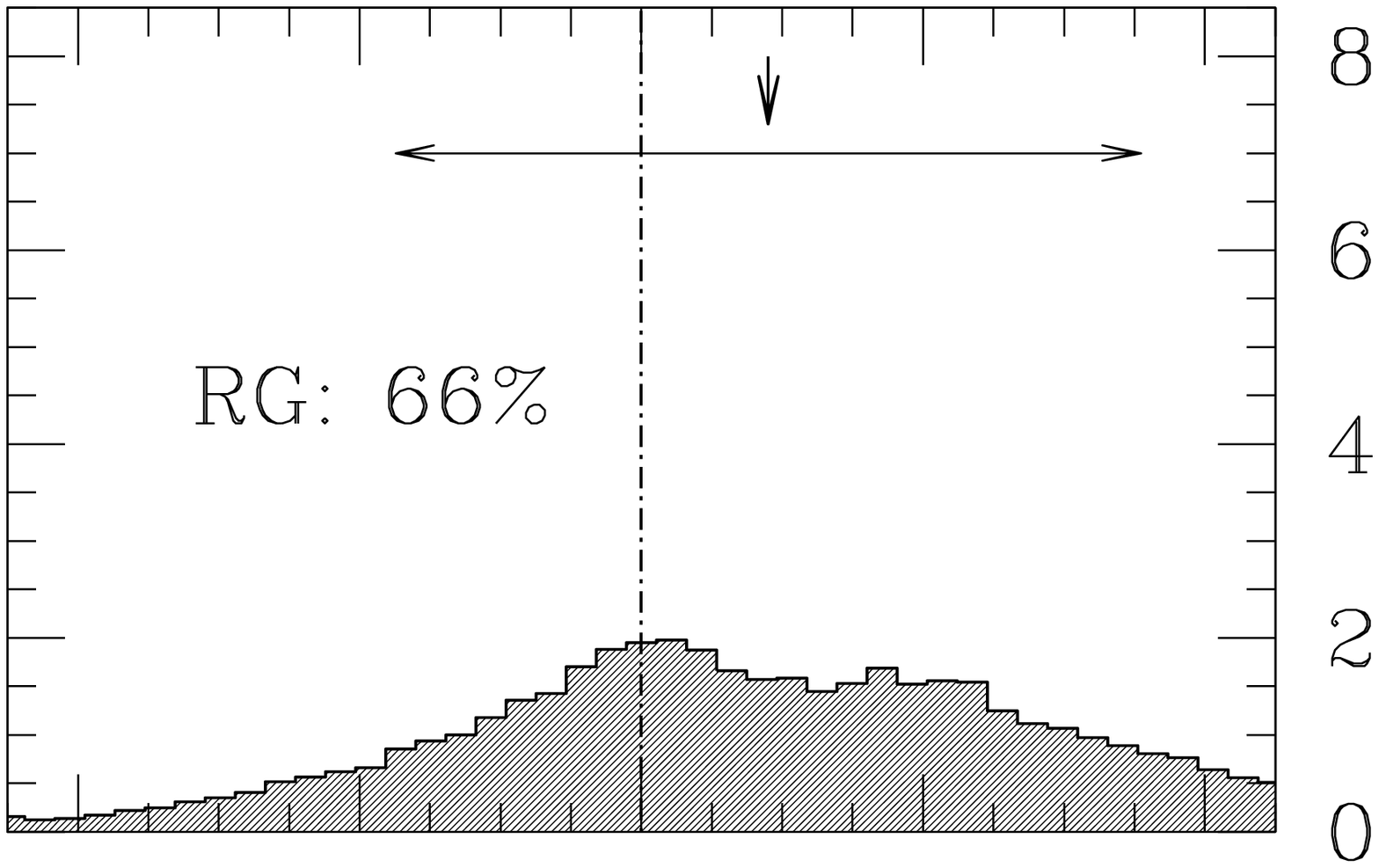}}
\centerline{\includegraphics[width=0.35\textwidth]{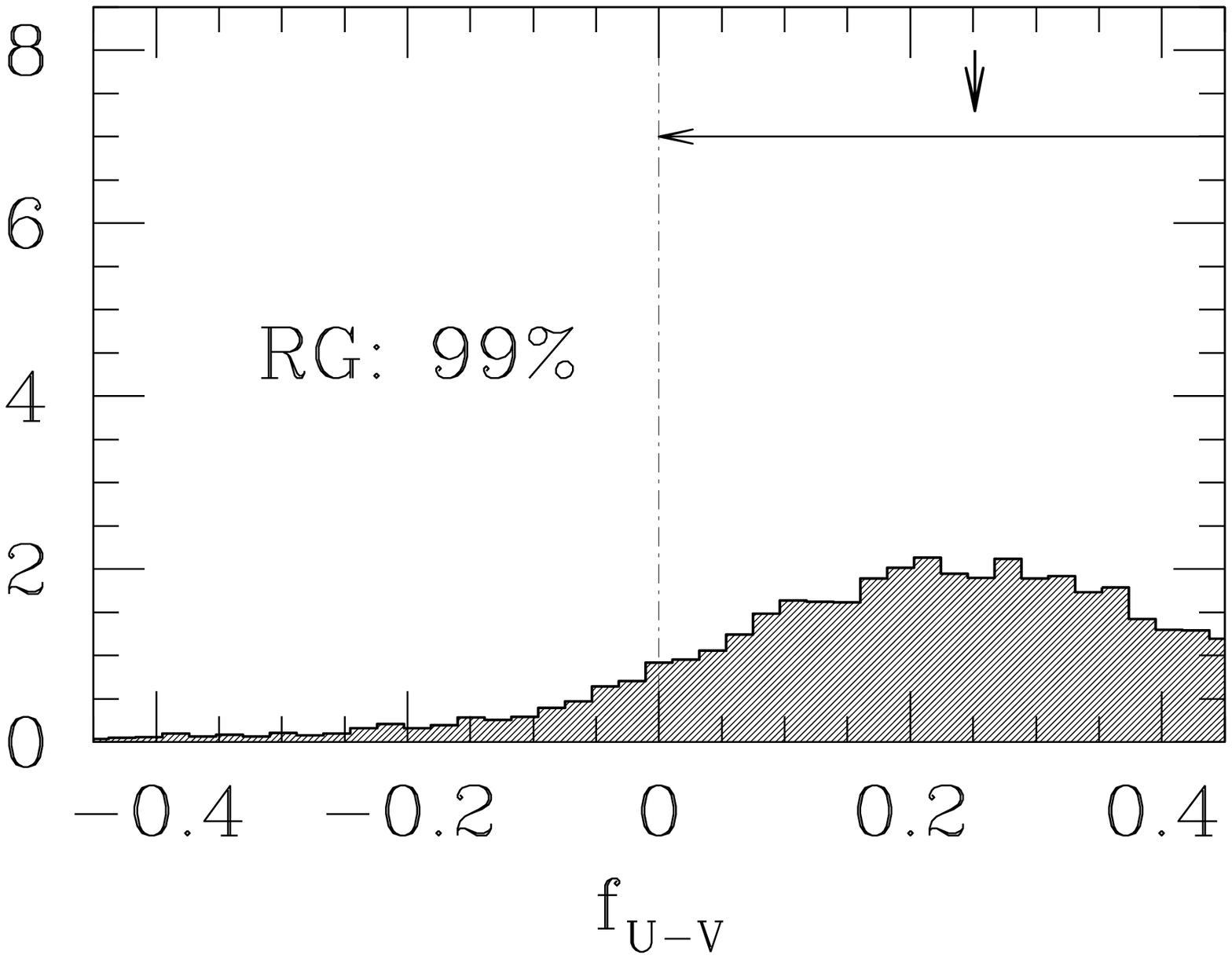}\includegraphics[width=0.35\textwidth]{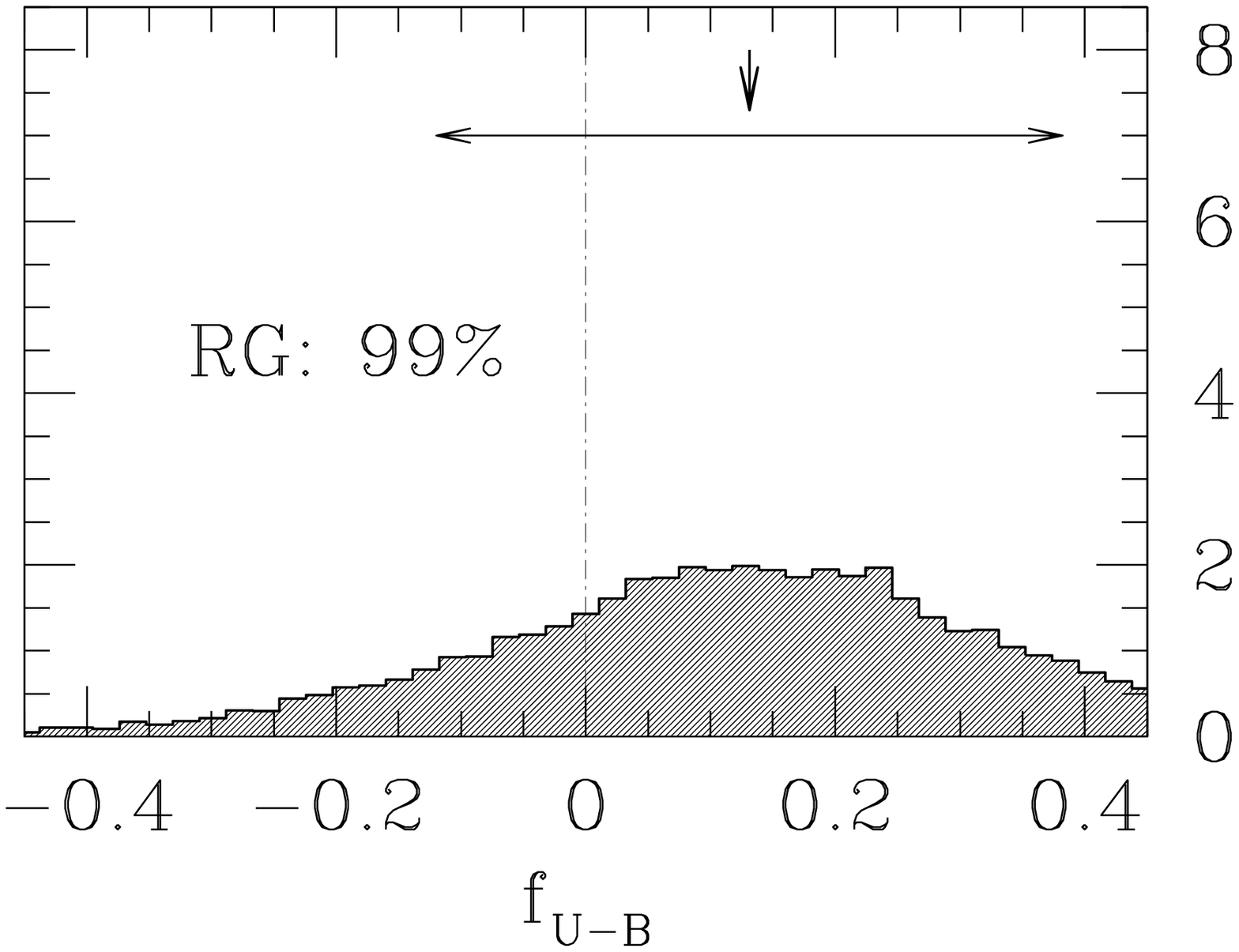}}
\caption{\UmV (left column) and \UmB (right column) distributions for true (top) and simulated data (four
 bottom plots) with different contribution of RG progenitors. 
The mean and standard deviation of each population are indicated by
 arrows. This
 Figure reproduces Figure~\ref{fig:BmV} for the \UmV and \UmB colors.}\label{fig:UmV}
\end{figure*}

 We limit ourselves to point
out that the \URF band data confirms the constraints that we set with
\BRF and \VRF data.

\section{Conclusions}\label{sec:conc}

We analyzed 3 years of spectroscopically confirmed \SNIae~from the SNLS survey looking for an early rise flux excess that could be attributed to shocking by a companion. We created composite lightcurve standardizing the data with the SiFTO method, excluding from the fit the region that might be affected by the shocking phenomenon.

We found a  worsening of the fit  of the data to the standard templates in the first few days after explosion (Section~\ref{sec:gof}), but we found no evidence that is due to anything but the fact that the data prior to $\trf = -10$ are not used in the template fit. 

We used the spectra generated by the K10 (Kasen 2010) simulations to model the expected \SNIa~time domain behavior in the SD scenario, thus we can account for sources of noise in the models, as well as in the data.
We found no evidence of flux excess in our data, and conclude that,
based on the K10 models, the contribution from RG progenitors is less than 10\% in the SNLS 3-year sample.
We thus set a $\sim 2\sigma$ upper limit of 10\% to the contribution of RG-WD binary
systems to the \SNIa~progenitors, and a $3\sigma$ upper limit of 20\%. With roughly 100 lightcurves in our
sample, with a contribution of $\sim 10\%$ lightcurves from RG progenitors, $\sim 3$
data points could be affected by shocking. We cannot exclude such a small
contribution from RG binary systems in the presence of noise from both the
models and the data. Our results are robust when tested in a
photometrically selected sample of lightcurves, as well as using \URF
band data.

Our conclusion agrees with the results derived in~\citet{2010sdss} from
the SDSS-II \SNIa~sample. Our analysis differs, other than in the SN
sample used, in the treatment of the shocking signature:
while~\citet{2010sdss} models the shocking as a Gaussian excess we used
the K10 simulation directly to characterize effects of shocking, thus
including the uncertainties in the models. Furthermore, our analysis
exploited the color bias in the shocking excess to set stronger
constraints on the presence of shocking and are able to quantify the
maximum allowed contribution of RGs to the \SNIa~progenitors.

Although Bayesian tests (Expectation Minimization and Gibbs sampling) were applied to our data, the presence of noise, and the relatively high dimensionality of the problem, with four possible progenitor scenarios, does not allow us to firmly asses what contribution of RGs best reproduces the residuals we see in the data with respect to the parabolic templates. Our data is entirely consistent with no RG progenitors.
                    
According to population synthesis studies, such as for
example~\citet{Ruiter09}, the SD scenario is expected to produce \SNIa~%
mainly from evolved companions, i.e. they favor the RG-WD channel over
the MS-WD channel, at least under the Roche lobe overflow
requirement. In \citet{Ruiter09}, where reaching the Chandrasekhar limit is
required, as it is assumed in this paper and in the K10 simulations,
MS-WD binaries are responsible only for 5\% to 10\% of the \SNIa~
production, while the majority of \SNIae~ come from a system with an evolved donor: a
sub-giant, or giant. Limiting the contribution of RG-WD
\SNIa~progenitors from an observational point of view may then have a
significant impact on the conclusions derived from population synthesis
stidies on delay time distributions and \SNIa~progenitors.

The PanSTARRS Medium Deep Survey (PS1, \citealt{2010ApJ...724L..16P})
and the Palomar Transient Factory (PTF, \citealt{2009PASP..121.1395L})
have begun providing well-sampled rise lightcurves, where the excess due
to shocking by a RG progenitor, should this be a valuable channel to
produce \SNIa, could soon be observed. Since the effect is predicted to
be chromatically biased (see Section~\ref{sec:models}), PS1 is
particularly suitable, offering data in SDSS \emph{g} and \emph{r}
bands, thus allowing a color comparison. Early UV follow up
surveys~\citep{2011ApJ...727L..35C} are also a promising way to spot
WD-RG progenitor systems, since the progenitor excess is extremely
prominent in UV bands, provided that the follow up can be triggered
early enough after explosion. SNLS continued collecting \SNIa~time
series through 2006 and as all SNLS data become available more stringent
limits may be set.  Note that in absence of any excess the population of
progenitors could be pinpointed to small ($M~<~ 6~M_\odot$) progenitor
companions. However, in the presence of detections of small excess
signals there would be a degeneracy between RG progenitors observed at
some angular offset from the line of sight to the hole generated by the
companion, and more massive MS companions, and a large sample is indeed
necessary to disentangle these two scenarios. Possibly, only a survey as
large as LSST \citep{LSST} would offer the opportunity to asses the
frequency of progenitor companion types in \SNIae.

We also point out that the constraints derived here rely on the
theoretical models described in K10. 
The shocking signatures predicted in K10 assume the
companion is in Roche lobe overflow, with the separation distance, $a$,
only a few times the stellar radius, $R$.  While this is expected in a
typical accretion scenario, if $a >> R$ the solid angle subtended by the
companion would be smaller, and so would be the effect of
shocking. \citet{2011apj...730L..34J}, for example, argues that the
donor star in the SD scenario might shrink rapidly before explosion,
having exhausted its envelope; the companion star would then be many
times smaller then its Roche lobe, reducing the shocking signature, and
also explaining the lack of hydrogen in spectra of \SNIae.
We look forward to
more detailed theoretical work, which may relax the Roche lobe overflow
assumption, integrate three dimensional explosion models, and takes into
account possible absorption mechanisms within the systems, and the
effects of the orbital motion, to better characterize the shocking
behavior and its diversity.

The authors wish to thank Lars Bildsten (KITP) and Ryan Foley (CfA) for
stimulating discussions and insightful comments. 

The SNLS collaboration gratefully acknowledges the assistance of Pierre
Martin and the CFHT Queued Service Observations team. Jean-Charles
Cuillandre and Kanoa Withington were also indispensable in making
possible real-time data reduction at CFHT. This paper is based in part
on observations obtained with MegaPrime/MegaCam, a joint project of CFHT
and CEA/DAPNIA, at the Canada-France-Hawaii Telescope (CFHT) which is
operated by the National Research Council (NRC) of Canada, the Institut
National des Sciences de l'Univers of the Centre National de la
Recherche Scientifique (CNRS) of France, and the University of
Hawaii. This work is based in part on data products produced at the
Canadian Astronomy Data Centre as part of the CFHT Legacy Survey, a
collaborative project of NRC and CNRS. MS acknowledges support from the
Royal Society. Canadian collaboration members acknowledge support from
NSERC and CIAR; French collaboration members from CNRS/IN2P3, CNRS/INSU
and CEA. Based in part on observations obtained at the Gemini
Observatory, which is operated by the Association of Universities for
Research in Astronomy, Inc., under a cooperative agreement with the NSF
on behalf of the Gemini partnership: the National Science Foundation
(United States), the Science and Technology Facilities Council (United
Kingdom), the National Research Council (Canada), CONICYT (Chile), the
Australian Research Council (Australia), CNPq (Brazil) and CONICET
(Argentina). Based on data from Gemini program IDs: GS-2003B-Q-8,
GN-2003B-Q-9, GS-2004A-Q-11, GN-2004A-Q-19, GS-2004B-Q-31,
GN-2004B-Q-16, GS-2005A-Q-11, GN-2005A-Q-11, GS-2005B-Q-6, GN-2005B-Q-
7, GN-2006A-Q-7, and GN-2006B-Q-10. Based in part on observations made
with ESO Telescopes at the Paranal Observatory under program IDs
171.A-0486 and 176.A-0589. Some of the data presented herein were
obtained at the W.M. Keck Observatory, which is operated as a scientific
partnership among the California Institute of Technology, the University
of California and the National Aeronautics and Space Administration. The
Observatory was made possible by the generous financial support of the
W.M. Keck Foundation.

\emph{Facilities}: CFHT, VLT:Antu, VLT:Kueyen, Gemini:Gillett, Gemini:South, Keck:I.


\end{document}